\documentclass[aps,superscriptaddress,floatfix]{revtex4}

\usepackage{amsmath}
\usepackage{amssymb}
\usepackage{graphicx}
\usepackage{slashed}
\usepackage{dsfont}
\usepackage{epsfig}
\usepackage{bm}

\begin{document}

\title{Information on the structure of the $a_1$ from $\tau$ decay}
\author{M. Wagner}
\affiliation{Institut f\"ur Theoretische Physik, Universit\"at Giessen, Germany}
\author{S. Leupold}
\affiliation{Institut f\"ur Theoretische Physik, Universit\"at Giessen, Germany}
\affiliation{Gesellschaft f\"ur Schwerionenforschung, Darmstadt, Germany}
\date{\today}

\begin{abstract}
The decay $\tau\rightarrow \pi\pi\pi\nu$ is analysed using different methods to account for the resonance structure, which is usually ascribed to the $a_1$. One scenario is based on the recently developed techniques to generate axial-vector resonances dynamically, whereas in a second calculation the $a_1$ is introduced as an explicit resonance. We investigate the influence of different assumptions on the result. In the molecule scenario the spectral function is described surprisingly well by adjusting only one free parameter. This result can be systematically improved by adding higher order corrections to the iterated Weinberg-Tomozawa interaction. Treating the $a_1$ as an explicit resonance on the other hand leads to peculiar properties. 
\end{abstract}

\maketitle

\section{Introduction}

\begin{figure}
\begin{center}
\begin{tabular}{cc}
\includegraphics[width=5.5cm]{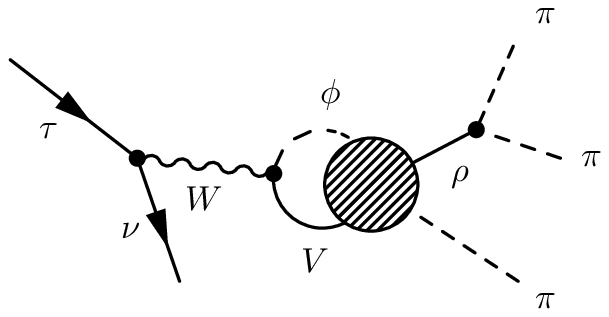} & \includegraphics[width=6cm]{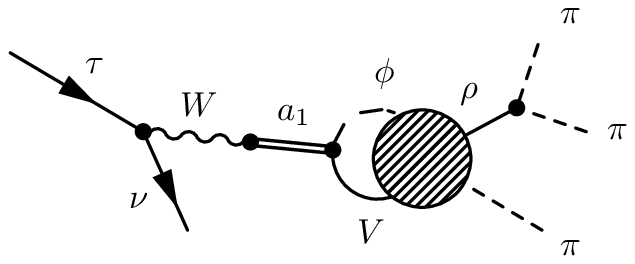}\\
(a) & (b) 
\end{tabular}
\caption{(a) Basic diagram describing the dynamically generated $a_1$ in the $\tau$ decay and (b) additional diagram, when the $a_1$ is included explicitly. $\phi$ and $V$ are the intermediate pseudoscalar and vector meson, respectively, which can be either $\pi\rho$ or $KK^\ast$.}\label{picintro}
\end{center}
\end{figure}

The constituent quark model \cite{godfrey,capstick,pdg} has been very successful in describing part of the observed hadron spectrum, especially for the heavy-quark systems, e.g. charmonia and bottomonia \cite{swanson}. On the other hand, especially in the light-quark sector, there is still a lively debate about the nature of many hadronic states. One sector with a lot of activity is, for example, the light scalar meson sector ($\sigma, a_0(980), f_0(980),\kappa(900)$). These states can not be explained within the naive constituent quark model, and many models have been proposed to explain the phenomenology of these resonances. The suggestions for the nature of these resonances vary between $q\overline q$ states, multiquark states, $K\overline K$ bound states and superpositions of them (see e.g. \cite{pdg,quarkrev,quarkrev2} and references therein). A different route to explain the low-lying scalars has been taken in \cite{osetscalar,nod} (see also references therein). In these works the authors explain the states as being dynamically generated by the interactions of the pseudoscalar mesons. The scattering amplitudes are calculated by iterating the lowest-order amplitudes of chiral perturbation theory (CHPT) \cite{scherer,gl84,gl2}, which leads to a unitarisation of the amplitudes and creates poles which can be associated with the scalars.\\
A similar question about the nature of hadronic resonances one encounters in the baryon sector, where the quark model also has trouble to describe the baryon excitations and their properties in a satisfying way (see e.g. \cite{baryon,baryon2} and references therein). As in the scalar case, an alternative approach to explain the resonance structure has been to generate resonances by iterating the leading order interactions of a chiral effective theory. The pioneering work in that direction has been done in \cite{weise1,weise2} and was followed by many other works \cite{bardyn1,bardyn2,bardyn4,lutz1,garcia}, which suggest a number of $J^P=\frac{1}{2}^-$ baryon resonances to be generated dynamically by the interactions of Goldstone bosons and baryons, e.g. $\Lambda(1405)$ and $N^\ast(1535)$. Studying the interaction of the pseudoscalar mesons with the decuplet of baryons \cite{bardyn3,bardyn5} also led to the generation of many known $J^P=\frac{3}{2}^-$ resonances, as e.g. the $\Lambda(1520)$.\\
Recent works applied the approach to the interactions of the octet of Goldstone bosons with the nonet of vector mesons focusing on the $J^P=1^+$ sector \cite{lutz2,osetaxial}. The authors calculate the scattering amplitude by solving a Bethe-Salpeter equation with a kernel fixed by the lowest-order interaction of a chiral expansion. The leading-order expression for the scattering of Goldstone bosons off vector mesons in a chiral framework is given by the Weinberg-Tomozawa (WT) term \cite{wt1,wt2} and leads to a parameter free interaction. The only free parameter in the calculation enters through the regularisation of the loop integral in the Bethe-Salpeter equation. Poles have been found, which have been attributed to the axial-vector mesons.\\
A comparison of the pole position and width is necessarily indirect and depends on the model, which is used to extract these quantities from the actual observables. In addition, the height of the scattering amplitude, or in other words the strength of the interaction, is not tested in this way. In the following we apply the method of dynamical generation directly to a physical process, namely the $\tau$ decay. The $\tau$ decay offers a clean probe to study the hadronic interactions since the weak interaction part is well understood and can be cleanly separated from the hadronic part, which we are interested in. The $\tau$ decay into three pions is dominated by a resonance structure, which is usually ascribed to the $a_1$ (see \cite{pdg} and references therein). Many of the references in \cite{pdg} are based upon a parametrisation in terms of Breit-Wigner functions, which leads to model dependent results. The relation of our calculation to some more microscopic descriptions \cite{pich,linsig} is discussed below.\\
We calculate the $\tau$ decay in two different ways: We first calculate it by assuming that the $a_1$ is generated dynamically and use the method from \cite{lutz2,osetaxial} to describe the decay ('molecule scenario'). This means that in this framework the $\tau$ decay is essentially described as follows: From the weak interactions a pair of mesons emerges (one pseudoscalar meson, one vector meson). Their final state interaction produces the resonant $a_1$ structure. This process is depicted in Fig. \ref{picintro}(a), where the blob stands for the iterated loop diagrams.
There are at most two free parameters in that calculation (in the simplest scenario only one), which enter in the renormalisation of the divergent loop integrals. All other parameters are fixed by chiral symmetry breaking and the properties of the $\rho$. In a second calculation, we introduce the $a_1$ explicitly. Here the idea is that the $a_1$ is a quark-antiquark state. At the hadronic level this substructure is not resolved and the $a_1$ should be included as an elementary field. A similar approach using chiral effective field theory including elementary vector mesons and axial-vector mesons has been performed in \cite{pich}. This method yields a good description of the spectral function for the decay into three pions. However, the width of the $a_1$ in \cite{pich} has been parametrised, whereas we generate the width by the $a_1$ decay into Goldstone bosons and vector mesons. In \cite{linsig} the authors successfully describe the spectral function for the decay $\tau^-\rightarrow 2\pi^0\pi^-\nu$ in the framework of the linear $\sigma$-model. The width of the $a_1$ in this model is generated from the elementary decays of the $a_1$. Nonetheless, there is still a fundamental difference between the approach we suggest in the present paper and the works \cite{pich,linsig}: Also for our second calculation with an elementary $a_1$ we still include the WT term since there is no reason to neglect it. The essential additional diagram is shown in Fig. \ref{picintro}(b), where the blob again represents the iterated loop diagrams, but that time the kernel also includes the $a_1$ interaction, which is discussed in detail in Section \ref{secdec}. Having both calculations at hand - one with and one without an elementary $a_1$ - we can compare both to experiment and see which scenario is favoured by the data. Since there exist excellent data for the $\tau$ decay \cite{aleph1}, one can expect that the results will be quite decisive. In case that the first scenario is favoured by experiment, this would be a sign that the $a_1$ is a dynamically generated resonance (molecule state) and in case the second calculation is favoured, this would be a hint that the $a_1$ is a quark-antiquark state. Some of the results of these calculations have already been shown in \cite{unsers}. In the present paper we present much more details of the calculations and show additional results, as for example the investigation of Dalitz plot projection data from \cite{dalitz} within the molecule scenario.\\
The $a_1$ is especially interesting, since it is considered to be the chiral partner of the $\rho$ \cite{weinberg2,scherer}. One expects a chiral partner for every particle from chiral symmetry. Due to the spontaneous symmetry breaking, one does not find degenerate one-particle states with the right quantum numbers. Nevertheless, the chiral partners have to exist, not necessarily as one-particle states, but at least as multi-particle states. Unmasking the $a_1$ as a bound state of a vector meson with a Goldstone boson would therefore approve its role of the chiral partner and disapprove its existence as a one-particle state. In the meson-meson and meson-baryon scattering examples, mentioned before, one can also see that some of the dynamically generated resonances would qualify as the chiral partners of the scattered particles, although the question of the chiral partner for these particles is not as clear as for the $a_1$ and the $\rho$. Even for the chiral partner of the $\rho$ a different suggestion besides the $a_1$ exists, namely the $b_1(1235)$ \cite{caldi}.\\
The work is structured as follows: We first discuss the general framework of our calculations in Section \ref{secinter}, and we write down the relevant interaction terms, which we will need during the calculation. Next we describe the unitarisation procedure to account for the final state correlations in Section \ref{secunit}. In Section \ref{secdec} we calculate the matrix elements for the $\tau$ decay in the different scenarios. Afterwards in Section \ref{secres} we compare our results to experiment and in Section \ref{secsum} we give a summary and an outlook. Further details on the formalism can be found in the appendix.
\section{Chiral interactions at tree level}\label{secinter}

The low-energy dynamics of the Goldstone bosons is described by chiral perturbation theory (CHPT) \cite{gl84,gl2,scherer}. To lowest order the Lagrangian reads
\begin{equation}\label{chirlag}
\mathcal L_\text{2}=\frac{F_0^2}{4}\text{Tr}[D_\mu U (D^\mu U)^\dagger] + \frac{F_0^2}{4}\text{Tr}[\chi U^\dagger + U \chi^\dagger]
\end{equation}
with
\begin{equation}
U=e^{i\phi/F_0}\,,\quad \chi=2B_0(s+ip)
\end{equation}
where
\begin{equation}
\phi=\begin{pmatrix}
\pi^0 +\frac{1}{\sqrt 3}\eta & \sqrt 2 \pi^+ & \sqrt 2 K^+ \\
\sqrt 2 \pi^- & -\pi^0+\frac{1}{\sqrt 3} \eta & \sqrt 2 K^0 \\
\sqrt 2 K^- & \sqrt 2\,\overline K^0 & -\frac{2}{\sqrt 3}\eta
\end{pmatrix}
\end{equation}
and the covariant derivative
\begin{equation}\label{covder}
D_\mu U=\partial_\mu U -ir_\mu U + i Ul_\mu\,.
\end{equation}
$r_\mu, l_\mu, s$ and $p$ are external fields, which promote the global SU(3)$\times$SU(3) symmetry to a local one. The interaction of the weak gauge boson with the Goldstone bosons can be determined by setting (see e.g. \cite{scherer})
\begin{equation}
r_\mu = 0\,,\quad l_\mu = -\frac{g}{\sqrt 2}(W^+_\mu T_+ + h.c.)\,,
\end{equation}
where $h.c.$ refers to the hermitian conjugate and
\begin{equation}
T_+ = \begin{pmatrix}
0 & V_{ud} & V_{us} \\
0 & 0 & 0 \\
0 & 0 & 0
\end{pmatrix}\,.
\end{equation}
The scalar field $s$ incorporates the explicit chiral symmetry breaking through the quark mass matrix. In the following it is sufficient to use $s=\text{diag}(m_u,m_d,m_s)$, $F_0$ is the pion decay constant in the chiral limit and $B_0$ parametrises the connection between the quark masses and the Goldstone boson masses. Following \cite{lutz2} we use $F_0=90\,$MeV throughout this work.\\
The Lagrangian Eq.(\ref{chirlag}) contains the direct coupling of the weak current to three pions, which describe the $\tau$ decay at very low energies \cite{Colangelo:1996hs}. At higher energies, however, the $\tau$ decay is dominated by resonances, most notably the vector mesons. In the molecule scenario we assume that the process is driven by the decay into Goldstone boson and vector meson. The structure, which is usually attributed to the $a_1$, is generated by the strong final state interactions of the Goldstone bosons and the vector mesons. This means that we need the interactions of the vector mesons with the Goldstone bosons and with the $W$ boson.

\subsection{Vector meson couplings}
Several models have been proposed to introduce the vector mesons in the chiral Lagrangian, e.g. the Hidden Symmetry approach \cite{bando} or the WCCWZ \cite{Weinberg:1968de,wzc,wzc2} scheme. Most of them were motivated by the phenomenological successful ideas of vector-meson dominance and universal coupling. We will use the WCCWZ scheme, where these features are implemented by putting constraints on the couplings. Besides the choice of the scheme to introduce the vector mesons, one also needs to choose the interpolating fields for the vector mesons. Instead of describing the particles in terms of four-vectors, the vector mesons can also be represented by antisymmetric tensor fields \cite{tenrep,vecrep}. The approaches are of course equivalent since the choice of fields can not influence the physics. However, due to the truncation in momentum the two descriptions can differ by higher order contact terms (see e.g. \cite{vecrep,Bijnens:1995ii}), which especially influence the behaviour at higher energies. In this work we use the WCCWZ scheme and describe the vector mesons by vector fields. In addition, we include contact terms to improve the high-energy behaviour.\\
The octet of vector mesons is given by
\begin{equation}
V_\mu = 
\begin{pmatrix}
\rho_\mu^0 + \omega_\mu^8/\sqrt 3  & \sqrt 2\rho_\mu^+ & \sqrt 2  K_\mu^+ \\ 
\sqrt 2 \rho_\mu^- & -\rho^0_\mu + \omega_\mu^8/\sqrt 3 & \sqrt 2 K_\mu^0 \\ 
\sqrt 2 K_\mu^- & \sqrt 2 \,\overline K_\mu^0 & -2 \omega_\mu^8/\sqrt 3 
\end{pmatrix}\,.
\end{equation} 
$\omega_\mu^8$ is an admixture of the physical states $\omega_\mu$ and $\phi_\mu$ (for details see e.g. \cite{pdg}). We do not care about the details of this mixing, since these states do not contribute to our calculation.\\
One can define a convenient representation by introducing the auxiliary quantity $u$, which is the square root of $U$ \cite{scherer} 
\begin{equation}
u^2=U\,.
\end{equation}
The transformation on $U$ under the chiral group induces a transformation on $u$, which is given by
\begin{equation}\label{kdef}
u\longrightarrow u^\prime = \sqrt{RUL^\dagger}\equiv RUK^{-1}(L,R,U) \,.
\end{equation}
The transformation of the vector fields under the chiral group in terms of $K$ is given by
\begin{equation}
V_\mu \rightarrow V^\prime_\mu = K(L,R,U)V_\mu K^\dagger (L,R,U) \,.
\end{equation}
The SU(3) matrix $K$ carries the $\text{SU(3)}_L\times\text{SU(3)}_R$ transformation in a non-linear way. The covariant derivative is defined for any object $X$, which transforms as $V_\mu$ 
\begin{equation}
\nabla_\mu X=\partial_\mu X +[\Gamma_\mu,X] \,,\quad \Gamma_\mu=\frac{1}{2}(u^\dagger(\partial_\mu-ir_\mu)u+u(\partial_\mu -il_\mu)u^\dagger))\,.
\end{equation}
In the following we also need $V_{\mu\nu}$, the field strength tensor of the vector mesons, which is given by
\begin{equation}
V_{\mu\nu} = \nabla_\mu V_\nu - \nabla_\nu V_\mu\,.
\end{equation}
Using vector fields in the above representation the already mentioned Weinberg-Tomozawa (WT) term is contained in the kinetic part of the Lagrangian. Together with the remaining relevant couplings of the vector mesons to lowest order \cite{vecrep}, one is led to the following interaction terms
\begin{equation}\label{lagvec}
\mathcal L_{vec} = -\frac{1}{2}\text{Tr}[[V^\nu,\partial_\mu V_\nu]\Gamma^\mu] -\frac{f_V}{4} \text{Tr}[V_{\mu\nu}f_+^{\mu\nu}] - \frac{ig_V}{4} \text{Tr}[V_{\mu\nu}[u^\mu,u^\nu]]\,,
\end{equation}
with
\begin{equation*}
u_\mu = i[u^\dagger(\partial_\mu-ir_\mu)u - u(\partial_\mu-il_\mu)u^\dagger],
\end{equation*}
and
\begin{equation*}
f^{\mu\nu}_{\pm}=u F_L^{\mu\nu} u^\dagger \pm u^\dagger F_R^{\mu\nu} u\,,
\end{equation*}
where $F_{L/R}^{\mu\nu}$ are the field strength tensors of the external left- and right-handed vector fields
\begin{equation}
F_{L}^{\mu\nu}= \partial^\mu l^\nu-\partial^\nu l^\mu-i[l^\mu,l^\nu]\,,\quad F_R^{\mu\nu}= \partial^\mu r^\nu-\partial^\nu r^\mu-i[r^\mu,r^\nu]\,.
\end{equation}
The first term in Eq.(\ref{lagvec}) is the WT term, which is parameter free. In a heavy-vector formalism \cite{lutz2,jenkins} it is the only term of order $\mathcal O(q^1)$, where $q$ is the momentum of the Goldstone bosons in a chiral counting. The Lagrangian in Eq.(\ref{lagvec}) has already been written down in \cite{vecrep}. We note that the definition of $V_\mu$ in \cite{vecrep} differs from our definition by a factor of $\sqrt 2$, which yields different coefficients in front of our terms. The two parameters $f_V,g_V$ can be determined from the decay of the $\rho$ into dileptons and two pions, respectively \cite{vecrep}, which yields
\begin{equation}\label{fvexp}
f_V = \frac{0.154\,\text{GeV}}{M_\rho}\,,\quad g_V=\frac{0.069\,\text{GeV}}{M_\rho}\,.
\end{equation}
In \cite{vecrep} the authors also give a theoretical estimate for these parameters, which yields
\begin{equation}
f_V =\frac{ \sqrt 2 F_0}{M_\rho}\approx \frac{0.127\,\text{GeV}}{M_\rho}\,, \quad g_V = \frac{F_0}{\sqrt 2 M_\rho}\approx\frac{0.064\,\text{GeV}}{M_\rho}\,.
\end{equation}
These values slightly differ from the experimental values, and in Section \ref{secres} we will study the influence of this difference. It will turn out that the experimentally measured values Eq.(\ref{fvexp}) describe the data best.

Transforming the vector-field Lagrangian Eq.(\ref{lagvec}) into a Lagrangian employing tensor fields, one finds that one can account for the difference resulting from the choice of fields by adding the following term (see e.g. \cite{vecrep,Bijnens:1995ii,kampf,stefan})
\begin{equation}\label{ho}
\mathcal L_{ho} = -2\text{Tr}[j_{\mu\nu}j^{\mu\nu}]
\end{equation}
with
\begin{equation*}
j^{\mu\nu} = -\frac{f_V}{4}f_+^{\mu\nu} - \frac{ig_V}{4}[u^\mu,u^\nu].
\end{equation*}
Thus, the entire Lagrangian we use is
\begin{equation}
\mathcal L = \mathcal L_2 + \mathcal L_{vec} + \mathcal L_{ho}\,.
\end{equation}
We will study the importance of $\mathcal L_{ho}$ below in Section \ref{secres}.

\subsection{Axial-vector meson couplings}
In the second scenario, where we introduce the $a_1$ explicitly, we also need to add the coupling of the axial-vector mesons to the Lagrangian. The nonet of axial-vector mesons $A_\mu$ is given by \cite{osetaxial}
\begin{equation}
A_\mu=
\begin{pmatrix}
a_1^0 + f_1(1285) & \sqrt 2 a_1^+ & \sqrt 2 K_{1A}^+ \\
\sqrt 2 a_1^- & -a_1^0 + f_1(1285) & \sqrt 2 K_{1A}^0 \\
\sqrt 2 K_{1A}^- & \sqrt 2 K_{1A}^0 & \sqrt 2 f_1(1420)
\end{pmatrix}_\mu \,.
\end{equation}
The additional Lagrangian we are going to use to describe the interactions of the $a_1$ is
\begin{equation}\label{axlag}
\mathcal L_{axial} = -\frac{f_A}{4}\text{Tr}[A_{\mu\nu}f^{\mu\nu}_-] + i c_1\text{Tr}[V^{\mu\nu}[A_\mu ,u_\nu]] + i c_2\text{Tr}[A^{\mu\nu}[V_\mu, u_\nu]]\,,
\end{equation} 
where the first term incorporates the coupling of the $a_1$ to the $W$ and the last two terms describe the decay of the $a_1$ into Goldstone boson and vector meson with the unknown constants $c_1$ and $c_2$. The first term has again already been written down in \cite{vecrep}, whereas one can find different approaches in the literature in order to describe the $a_1$ decay vertex. In \cite{oseta1} the authors propose a phenomenological Lagrangian in terms of tensor fields, which is successful in reproducing the decay branching ratios. In \cite{meissnera1} the hidden symmetry formalism was used to derive the pertinent terms, which yields the same results as the phenomenological approach. Comparing the vertex resulting from the Lagrangian above to these works, we find agreement by choosing
\begin{equation}\label{c1c2}
c_1 = -\frac{1}{4}\,,\quad c_2=-\frac{1}{8}\,.
\end{equation}
Below we will also study variations of $c_1$ and $c_2$ around these values. Looking at Eq.(\ref{axlag}) we see that both terms describing the decay into vector and pseudoscalar meson contain one $u^\mu$. This means that integrating out the axial-vector fields would generate to lowest order an interaction term of vector mesons and Goldstone bosons, which contains two $u^\mu$ and therefore leads to an expression of order $\mathcal O(q^2)$. Since the WT term is $\mathcal O(q^1)$ including both interactions is not double counting. This will be an important aspect for our results presented below.

\section{Unitarisation procedure}\label{secunit}

\begin{figure}
\begin{center}
\includegraphics[width=12cm]{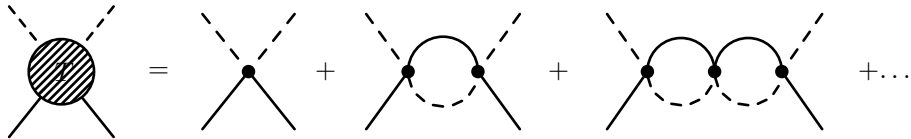}
\caption{Iteration of loop diagrams, corresponding to the approximation to the Bethe-Salpeter equation by using the WT term as kernel. Full lines denote vector mesons, dashed lines Goldstone bosons.}\label{bspic}
\end{center}
\end{figure}

\begin{figure}
\begin{center}
\epsfig{file=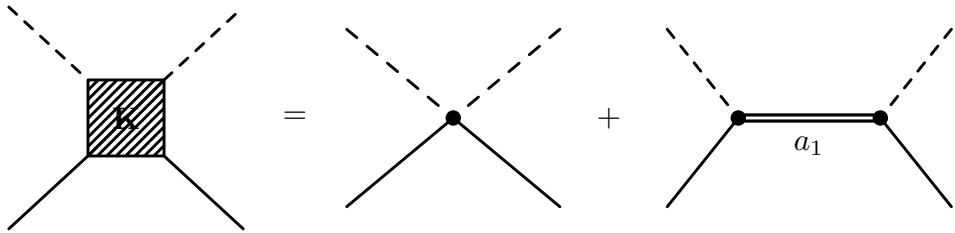, height=3cm}
\caption{Kernel of the Bethe-Salpeter equation when including the $a_1$ explicitly; i.e. one has to replace the point interaction in Fig. \ref{bspic} by the two diagrams on the right hand side.}\label{kerna1}
\end{center}
\end{figure}

Unitarisation methods have been used to extend the applicability of the chiral Lagrangians to higher energies, e.g. the N/D method \cite{nod}, the inverse amplitude method \cite{iam1} or partial summations using the Bethe-Salpeter equation \cite{lutz2,osetaxial}. In \cite{nod} it was shown that the N/D method is equivalent to a summation of diagrams using the Bethe-Salpeter equation with the kernel taken onshell. We follow the work in \cite{lutz2,osetaxial} and use the Bethe-Salpeter equation to describe the final state interactions between vector and pseudoscalar mesons. In the molecule scenario this serves to generate the axial-vector mesons dynamically. In the scenario with the explicit $a_1$ the width of the latter is generated. The iteration (of a point interaction) is shown diagrammatically in Fig. \ref{bspic}.
In case we are not including the $a_1$ explicitly, we will use the amplitudes from \cite{lutz2,osetaxial}, i.e. the iterated WT interaction, in order to describe the final state correlations of the vector meson and the Goldstone boson. In the scenario, where we explicitly take into account the $a_1$, we include, in addition to the WT term, the $a_1$ interaction (cf. Fig. \ref{kerna1}). 
This in principle generates the width of the $a_1$ by the decay into vector meson and Goldstone boson with the WT term as correction.\\
Concerning the choice for the kernel in the case that we include the $a_1$ explicitly, we note the following. The kernel of the Bethe-Salpeter equation can consist of parts, which are analytic in the energy region, one is interested in, and parts which are non-analytic. The non-analytic parts are possible $s$-channel resonances, which have a pole in the physical region. The analytic part consists of contact terms, as well as $t$- and $u$-channel processes, since they do not have any singularities in the physical energy region. The analytic parts can be expanded in powers of the momenta of the involved particles (and in powers of the Goldstone boson masses). This leads to contact terms, which can equivalently and systematically be expressed in terms of a chiral Lagrangian. The approach we take is to consider the relevant s-channel resonances in the kernel and keep contact terms up to a specific order. This approximation is a model assumption, and it is not guaranteed that it works for the quite large energy region, we are interested in. However, it is certainly worthwhile to study its properties. Following this strategy, one has to avoid double counting between the s-channel processes and the contact terms, since for $s<m_{res}^2$ these $s$-channel resonance terms can also contribute to the analytic part. Restricting the contact terms to the WT term, there is no problem in our case, since the WT term contributes at $\mathcal O(q)$ and the elementary $a_1$ at $\mathcal O(q^2)$. Thus, there can be no double counting. In particular, this means that only considering an $a_1$ and neglecting the WT term is very questionable. The focus in discussing the results will therefore be on the interplay between the $a_1$ and the WT term. In order to check the systematics of our model, we will also study the influence of keeping contact terms up to order $\mathcal O(q^2)$ instead of $\mathcal O(q)$ in the scenario without explicit $a_1$.\\
We will briefly summarise the formalism, which we employ to iterate the respective kernels, which basically follows the same lines as in \cite{lutz2}. The Bethe-Salpeter equation for the coupled channel problem is
\begin{equation}
T_{\mu\nu}^{ab}(\overline q,q,w)= K_{\mu\nu}^{ab}(\overline q,q,w) + \sum_{c,d} \int \frac{d^4 l}{(2\pi)^4} K_{\mu\beta}^{ad}(\overline q ,l,w) G^{\alpha\beta}_{dc}(l,w) T_{\alpha\nu}^{cb}(l,q,w)
\end{equation}
where $T$ is the scattering amplitude, $K$ the kernel, $q$ $(\overline q)$ is the incoming (outgoing) momentum of the Goldstone boson, $w$ the total four momentum and
\begin{equation}
G_{\alpha\beta}^{cd} = \frac{i}{(w-l)^2 - m_\phi^2 + i\epsilon} \frac{g_{\alpha\beta} - \frac{l_\alpha l_\beta}{M_V^2}}{l^2 - M_V^2 + i\epsilon}\delta_{cd}
\end{equation}
is the two-particle propagator. The indices $a,b,c,d$ indicate the channel, which in our case can be either $\pi\rho$ or $KK^\ast$. The indices are chosen such that $T_{\mu\nu}^{ab}$ denotes the scattering of $(b,\nu)\rightarrow(a,\mu)$.\\
We will solve the Bethe-Salpeter equation by employing an expansion in helicity amplitudes as follows
\begin{equation}
K_{ab}^{\mu\nu}(q,\overline q,w)=\sum_{J,M,P,i,j} V^{(JMP)}_{abij}(s) Y^{\mu\nu (JMP)}_{ij}(q,\overline q,w)\,,
\end{equation}
\begin{equation}
T_{ab}^{\mu\nu}(q,\overline q,w)=\sum_{J,M,P,i,j} M^{(JMP)}_{abij}(s) Y^{\mu\nu (JMP)}_{ij}(q,\overline q,w)\,,
\end{equation}
where $s=w^2$ is the invariant mass of the hadronic final state. The indices $i,j$ correspond to the helicities and can be either $0$ or $1$ in our calculation. The objects $Y^{\mu\nu (JMP)}_{ij}$ are similar to the projectors in \cite{lutz2} and are discussed in Appendix \ref{secproj}. In order to arrive at an expansion with coefficients which only depend on $s$, one has to take the amplitudes onshell. This has been discussed in several works and we refer to \cite{osetscalar,oseta1,lutz2} for a justification.\\
In \cite{lutz2} the authors were concerned that the projectors are not analytic outside the centre-of-mass system (CMS) and therefore performed a transformation to covariant projectors, which mix different helicities. We will not do that transformation since we only work in the CMS and without this transformation it will be easier to connect to states with definite angular momentum (see Appendix \ref{orbsec}). The projectors fulfil the following orthogonality relation
\begin{equation}
\int \frac{d^4 l}{(2\pi)^4} \, Y_{\lambda_1\lambda_2 \mu\alpha}^{JMP}(\overline q,l,w) G^{\alpha\beta}(l,w) Y_{\lambda_3 \lambda_4 \beta\nu}^{J^\prime M^\prime P^\prime}(l,q,w)=\delta_{\lambda_2\lambda_3}\delta_{PP^\prime}\delta_{JJ^\prime}\delta_{MM^\prime} Y_{\lambda_1\lambda_4 \mu\nu}^{JMP}(\overline q,q,w)(-I_{\phi V})\,,
\end{equation}
with the divergent loop integral
\begin{equation}
I_{\phi V}(s) = \int \frac{d^4 l}{(2\pi)^4}\frac{i}{(w-l)^2-m_\phi^2+i\epsilon}\frac{1}{l^2-M_V^2+i\epsilon}\,.
\end{equation}
The relation only holds up to additional tadpoles, which are dropped. Using this relation the Bethe-Salpeter equation turns into an algebraic equation for the expansion coefficients
\begin{equation}\label{bsalg}
M_{abij} = V_{abij} + \sum_c \sum_k  V_{acik} M_{cb kj} (-I_{\phi V})\,.
\end{equation}
We introduce the renormalised quantity $J_{\phi V}(s,\mu)$ 
\begin{equation}
J_{\phi V}(s,\mu) = I_{\phi V}(s) - I_{\phi V}(\mu)\,,
\end{equation}
which depends on the subtraction point $\mu$. In order to render Eq.(\ref{bsalg}) finite we substitute 
\begin{equation}
I_{\phi V}(s) \rightarrow J_{\phi V}(s,\mu_1)\,,
\end{equation}
which introduces the a priori unknown parameter $\mu_1$. It remains to determine the coefficients $V_{cbik}$ in order to calculate the scattering amplitude. The relevant formulas are given in Appendix \ref{secproj}.

\section{$\tau$ decay \label{secdec}}

\subsection{The decay width}\label{secwidth}

Since the weak decay vertex is common in all diagrams (cf. Fig. \ref{diags1} and Fig. \ref{diags2} below), we can separate it from the hadronic information by writing the invariant matrix element as
\begin{equation}\label{decomp}
i\mathcal M = C(s) S_\mu \left(g^{\mu\nu}-\frac{w^\mu w^\nu}{M_W^2}\right)W_\nu \,,
\end{equation}
where we used the following abbreviations
\begin{equation}\label{smu}
S_\mu=\overline v(p_\nu)\gamma_\mu(1-\gamma_5)u(p_\tau)
\end{equation} 
and
\begin{equation}\label{bigc}
C(s) = \left(\sqrt{\frac{G_F M_W^2}{\sqrt 2}} \right) \frac{1}{s-M_W^2} \simeq -\sqrt{\frac{G_F}{\sqrt 2 M_W^2}} = C \,.
\end{equation} 
$W_\nu$ denotes the hadronic tensor, which we will calculate in detail below. $G_F$ is the Fermi constant, which is connected to the weak gauge coupling by $G_F=\frac{g^2}{4\sqrt 2 M_W^2}$. The decay width is given by
\begin{equation}\label{decwidth}
d\Gamma = \frac{(2\pi)^4}{2M_\tau} |\mathcal M|^2 d\phi_4\,,
\end{equation}
where $d\phi_4$ is the four-body phase space. We define the hadronic tensor
\begin{equation}
W_{\mu\nu} \equiv \int_{\phi_3} W_\mu^\ast W_\nu d\phi_3\,,
\end{equation}
which by Lorentz invariance must have the following structure 
\begin{equation}\label{w1w2}
W^{\mu\nu}(s) = W_1(s)\left(g^{\mu\nu}-\frac{w^\mu w^\nu}{w^2}\right) + W_2(s)\frac{w^\mu w^\nu}{w^2}\,.
\end{equation}
Plugging in Eq.(\ref{decomp}) into Eq.(\ref{decwidth}) and expressing the width in terms of $W_1$ and $W_2$ we get for the differential decay width for the process $\tau^-\rightarrow 2\pi^0\pi^-\nu_\tau$
\begin{equation}\label{gamma}
\frac{d\Gamma}{ds} = \frac{\pi^2}{2M_\tau s}|C|^2(M_\tau^2-s)^2\left(W_2-W_1\left(1+\frac{2s}{M_\tau^2}\right)\right)\,.
\end{equation}
$W_1$ and $W_2$ can then be calculated from the following two equations
\begin{align}
3 W_1(s) + W_2(s) &= \int_{\phi_3} W^\ast\cdot W d\phi_3 \,, \label{w1}\\
W_2(s) &= \frac{1}{s}\int_{\phi_3} w\cdot W^\ast w\cdot W d\phi_3\,. \label{w2}
\end{align}
The factor of $\frac{1}{2}$ taking care of the identical particles in the final state has been introduced in Eq.(\ref{gamma}), and thus it does not appear in Eq.(\ref{w1}) and Eq.(\ref{w2}). The longitudinal part $W_2$ has no visible effect on the calculation, which was expected since it is proportional to $m_\pi^2$ as can be seen below. Therefore, we will drop it from now on.\\
The quantity, we compare most of our calculations with, is the spectral function $a_1(s)$ for the decay $\tau\rightarrow 2\pi^0\pi^-\nu_\tau$. The complete spectral function $A(s)$ is defined by
\begin{equation}
A(s) = -\frac{2\pi}{s} \Im \Pi_T(s)\,,
\end{equation}
with the hadronic vacuum polarisation
\begin{equation}
\Pi_{\mu\nu} = \Pi_T\left(g_{\mu\nu}-\frac{w_\mu w_\nu}{s}\right) + \Pi_L\frac{w_\mu w_\nu}{s}
\end{equation}
and
\begin{equation}
\Pi_{\mu\nu} = i \int d^4 x e^{iwx}\langle 0 | T A_\mu(x) A_\nu(0)^\dagger |0\rangle\,,
\end{equation}
where $T$ is the time ordering symbol and $A^\mu$ is the charged axial current
\begin{equation}
A_\mu = \overline u \gamma_\mu\gamma_5 d\,.
\end{equation}
In terms of $W_1$ the spectral function for the decay $\tau\rightarrow 2\pi^0\pi^-\nu_\tau$ is given by
\begin{equation}
a_1(s) = -\frac{2^6 \pi^5}{g^2 V_{ud}^2 s}W_1\,.
\end{equation}

\subsection{The general picture}\label{secwhich}

\begin{figure}
\begin{center}
\includegraphics[width=9cm]{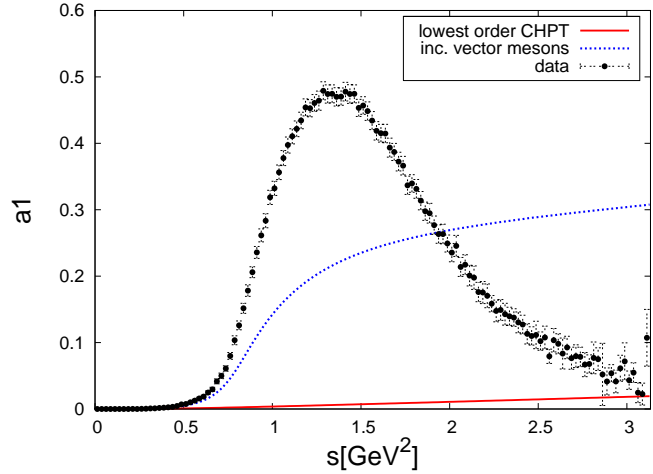}
\caption{Spectral function for the decay $\tau^- \rightarrow 2\pi^0\pi^-\nu$ without including rescattering diagrams in comparison to data from \cite{aleph1}. The lowest-order CHPT calculation corresponds to the diagrams Fig. \ref{diags1}a, \ref{diags1}b and the second curve ('inc. vector mesons') additionally includes the diagrams Fig. \ref{diags1}c, \ref{diags1}d.}\label{3pidir}
\end{center}
\end{figure}

\begin{figure}
\begin{center}
\begin{tabular}{cc}
\includegraphics[width=4.5cm]{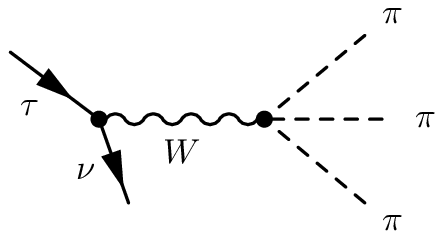} & \includegraphics[width=5cm]{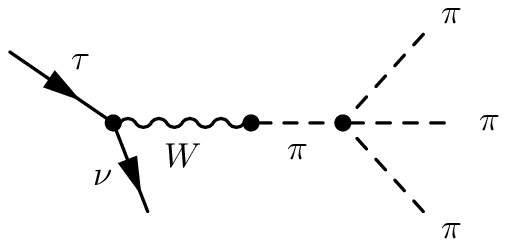} \\
(a) & (b) \\
\includegraphics[width=5cm]{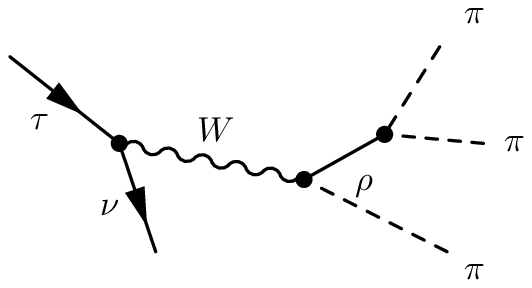} & \includegraphics[width=5.5cm]{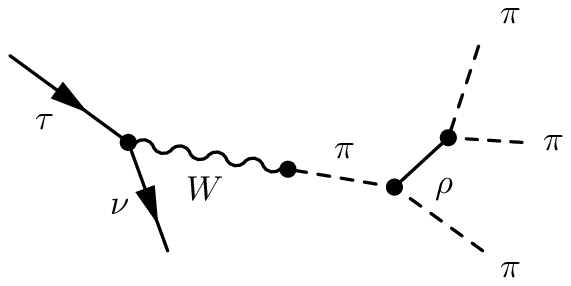} \\
(c) & (d) \\
\includegraphics[width=5.5cm]{taudecpic5} & \includegraphics[width=6cm]{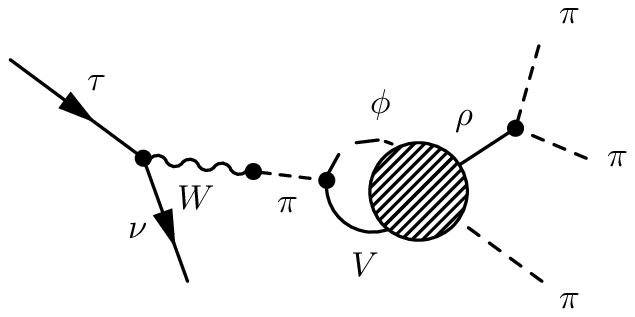} \\ 
(e) & (f) \\
\end{tabular}
\caption{Relevant diagrams for the decay $\tau^- \rightarrow 2\pi^0\pi^-\nu$ without including the $a_1$. $\phi$ and $V$ correspond to intermediate Goldstone boson and vector meson ($\pi\rho$ or $KK^\ast$). The blob represents the final state interactions obtained from the solution of the Bethe-Salpeter equation (see Fig. \ref{bspic}).}\label{diags1}
\end{center}
\end{figure}

\begin{figure}
\begin{center}
\includegraphics[width=4cm]{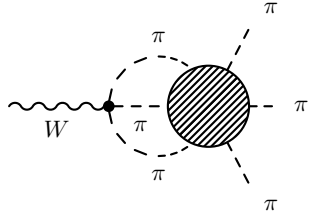}
\caption{Diagram describing pion correlations, which we do not include in our calculation. The blob denotes the final state interactions of the pions.}\label{notinc}
\end{center}
\end{figure}

One can not expect chiral perturbation theory to describe the $\tau$ decay in the whole energy region (see \cite{Colangelo:1996hs} for pure CHPT calculations), since the energies, which are involved are beyond 1$\,$GeV and the decay is dominated by resonance structures. Including vector mesons at tree level in the calculation will certainly improve the calculation, but still one can not expect to find a satisfying description of the data due to the strong correlations in the final state. In particular, the vector mesons at tree level can not produce an axial-vector resonance. In Fig. \ref{3pidir} we see the spectral function calculated in lowest-order CHPT and by including vector mesons in comparison to data. 
The lowest-order CHPT calculation using the Lagrangian Eq.(\ref{chirlag}) (cf. Fig. \ref{diags1}a, \ref{diags1}b) can only describe the lowest data points, which are far below the $\pi\rho$ threshold. The onset of the rise in the region $0.5-0.7\,$GeV$^2$ is described much better if one includes the tree-level vector-meson diagrams Fig. \ref{diags1}c, \ref{diags1}d. Nonetheless, the main bump in the data at about $1.5\,$GeV$^2$ is clearly out of reach. The philosophy in the present work is that the final state interaction is dominated by the coupled-channel dynamics of the $\pi\rho$ state. That point of view is suggested by the improvement when we include the $\rho$ in the calculation and the height of the amplitude, which we can see in Fig. \ref{3pidir}. The deviation from the data for higher energies can then be explained by the increasing importance of the rescattering diagrams, describing the final state interactions (Fig. \ref{diags1}e, \ref{diags1}f). We neglect further correlations between the pions, e.g. the diagram in Fig. \ref{notinc}, which we expect to have only a minor influence, since the tree-level result for the three-pion process is much smaller than the tree-level result for the process including vector mesons (see Fig. \ref{3pidir}).\\
When we include the $a_1$ we expect, of course, also the $a_1$ to have a major influence on the result, but we recall that we still include the WT term in the calculation.\\
In Section \ref{secres} we will also look at Dalitz plot projections, where one can see the clearly dominating $\rho$ in the final state, which gives another justification for the choice of processes, which we include.\\
In the present section we will present the formalism for different scenarios. Results are postponed to Section \ref{secres}.

\subsection{Calculation of $\tau$ decay without $a_1$}\label{secwwt}
In Fig. \ref{diags1} we see the processes which we take into account. The diagrams Fig. \ref{diags1}a, \ref{diags1}b are the lowest-order CHPT processes, Fig. \ref{diags1}c, \ref{diags1}d are the tree-level processes including vector mesons and the diagrams Fig. \ref{diags1}e, \ref{diags1}f describe the rescattering. $W_\mu$ from Eq.(\ref{decomp}) is split into these contributions
\begin{equation}\label{hadten}
W^\mu = W^\mu_{3\pi} + W_{vec}^\mu + W_{\pi\rho}^\mu + W_{KK^\ast}^\mu\,,
\end{equation}
where $W^\mu_{3\pi}$ corresponds to the processes in Fig. \ref{diags1}a, \ref{diags1}b, $W_{vec}^\mu$ to the diagrams Fig. \ref{diags1}c, \ref{diags1}d and $W_{\phi V}^\mu$ to Fig. \ref{diags1}e, \ref{diags1}f. The first two functions are given by
\begin{align}
W^\mu_{3\pi} &= -\left(g^{\mu\nu}-\frac{w^\mu w^\nu}{w^2}\right)\frac{gV_{ud}}{F_0}\, q_{2\nu} - \frac{gV_{ud}}{F_0} \frac{w^\mu}{s}\frac{m_\pi^2}{s-m_\pi^2}\left(\frac{1}{2}s - (q_2\cdot w)\right) \\
W^\mu_{vec} &= \left(g^{\mu\nu} -\frac{w^\mu w^\nu}{w^2}\right)\frac{gV_{ud}g_V}{ F_0^3}\frac{m_{12}^2}{m_{12}^2-M_\rho^2- \Pi}\Bigl(f_V(m_{12}^2 q_{2\nu} - (m_{12}\cdot q_2) m_{12\nu})\nonumber \\ 
& \phantom{(g^{\mu\nu} -\frac{w^\mu w^\nu}{w^2})\frac{gV_{ud}g_V}{ F_0^3}\frac{m_{12}^2}{m_{12}^2-M_\rho^2-\Pi}} + (f_V- 2g_V)((m_{12}\cdot q_3)q_{2\nu} -m_{12\nu} (q_3\cdot q_2))\Bigr) \nonumber \\
& + \frac{w^\mu}{s} \frac{m_\pi^2}{s-m_\pi^2}\frac{2gV_{ud}g_V^2}{ F_0^3}\bigl((m_{12}\cdot q_3)(w\cdot q_2) - (w\cdot m_{12})(q_3\cdot q_2)\bigr)\frac{m_{12}^2}{m_{12}^2-M_\rho^2 -\Pi}\quad +\quad  (q_1\leftrightarrow q_3) \label{thews3}\,,
\end{align}
where $q_1$ and $q_3$ are the momenta of the likewise non-charged pions, $q_2$ is the momentum of the charged pion, $w=q_1+q_2+q_3$, $m_{ij}= q_i + q_j$ and
the self energy of the vector mesons $\Pi$ is taken from \cite{spec}. $(q_1\leftrightarrow q_3)$ denotes the same amplitude with the pion momenta $q_1$ and $q_3$ exchanged, which arise due to the appearance of two identical pions. The rescattering part is given by 
\begin{equation}
W^\mu_T = W^\mu_{\pi\rho}+ W^\mu_{KK^\ast} = \left(g^{\mu\nu} - \frac{w^\mu w^\nu}{w^2}\right)b_{T}\frac{m_{12}^2}{m_{12}^2-M_\rho^2 -\Pi} (q_1-q_2)_\nu \quad + \quad (q_1 \leftrightarrow q_3)\,,
\end{equation}
where the nontrivial information is contained in $b_T$. $W_\mu^T$ contains an additional loop diagram, which needs to be renormalised. As in the Bethe-Salpeter equation \cite{lutz2} we also drop additional tadpoles here, which leads to
\begin{equation}
\begin{split}
&b_T = \frac{3 g_V g V_{ud}}{4 F_0^3}M_{1111}J_{\pi\rho}(\mu_2) \left((f_V-2g_V)\frac{1}{2}(s-m_\pi^2+M_\rho^2)+ 2g_V \left(\frac{2}{3}M_\rho^2 + \frac{1}{12s}(m_\pi^2-M_\rho^2-s)^2\right)\right)\\
& -\frac{3 g_V g V_{ud}}{4 \sqrt 2 F_0^3}M_{1211} J_{KK^\ast}(\mu_2)\left((f_V-2g_V)\frac{1}{2}(s-m_K^2+M_{K^\ast}^2) + 2g_V \left(\frac{2}{3}M_{K^\ast}^2 + \frac{1}{12s}(m_K^2-M_{K^\ast}^2-s)^2\right)\right)\,.
\end{split}
\end{equation}
The expansion coefficients for the scattering amplitude $M_{abij}$ have been given in Section \ref{secunit}. We numbered the different isospin channels, where the channel 1 corresponds to $\pi\rho$ and 2 denotes $KK^\ast$. Thus, e.g. $M_{1111}$ denotes the amplitude for $\pi\rho$ scattering with the helicity of the $\rho$ equal 1 in incoming and outgoing channel. We note that the renormalised loop integral $J_{\phi V}(\mu_2)$ appearing in $b_T$ does not have to depend on the same subtraction constant as the loop integrals in the scattering amplitude. Therefore we denote the subtraction constant of this loop with $\mu_2$ and the subtraction constant appearing in the scattering amplitude with $\mu_1$. We will discuss the appearance of two subtraction constants and their relation in more detail later.\\
The scattering amplitude we used for this calculation was already determined in \cite{lutz2} and \cite{osetaxial} and in terms of the expansion in the projectors from Appendix \ref{secproj} we have
\begin{align}
V_{ab11}^{1^+} &= g_{ab}\,, \\
V_{ab01}^{1^+} &= \frac{\omega_a}{\sqrt 2 M_{Va}} g_{ab}\,,\\
V_{ab10}^{1^+} &= \frac{\omega_b}{\sqrt 2 M_{Vb}} g_{ab}\,,\\
V_{ab00}^{1^+} &= \frac{\omega_a\omega_b}{2M_{Va} M_{Vb}} g_{ab} - \frac{C_{WTab} p^2_{cma} \overline p^2_{cmb}}{6F_0^2 M_{Va} M_{Vb}} \label{lastline}
\end{align}
with
\begin{equation}\label{k0}
g_{ab} = \frac{C_{WTab}}{12F_0^2}\left(3s - (M_{\phi a}^2 + M_{\phi b}^2 + M^2_{Va} + M_{Vb}^2) -\frac{1}{s}(M^2_{Vb} -M_{\phi b}^2)(M_{Va}^2 -M_{\phi a}^2)\right)\,,
\end{equation}
\begin{equation}
\omega_a = \frac{1}{2\sqrt s}(s+M_{Va}^2-M_{\phi a}^2)\,,
\end{equation}
\begin{equation}
p_{cma}=\frac{1}{2\sqrt s}\sqrt{(s-(M_{Va}+ M_{\phi a})^2)(s-(M_{Va}-M_{\phi a})^2)}
\end{equation}
and
\begin{equation}
C_{WT} = 
\begin{pmatrix}
2 & -\sqrt 2 \\
-\sqrt 2 & 1
\end{pmatrix}\,.
\end{equation}
The scattering amplitude is then determined by Eq.(\ref{bsalg}). The second term in Eq.(\ref{lastline}) is in principle a higher-order term and we neglect it for the moment. We explicitly checked the influence of this term and there was no visible difference in the results by including the term. With these coefficients the lowest partial wave of the potential takes an easy form
\begin{equation}\label{wtkern}
\begin{split}
K_{\mu\nu}^{J^P=1^+} &= g_{ab} Y_{11\mu\nu}^{1^+} + g_{ab}\frac{\omega_a}{\sqrt 2 M_{Va}} Y_{01\mu\nu}^{1^+} + g_{ab}\frac{\omega_b}{\sqrt 2 M_{Vb}} Y_{10\mu\nu}^{1^+} + g_{ab} \frac{\omega_a\omega_b}{2M_{Va}M_{Vb}}Y_{00\mu\nu}^{1^+}\\
&= -\frac{3}{2} g_{ab} L^1_{\mu\nu} = -g_{ab} \frac{3}{2} \left(g_{\mu\nu}-\frac{w_\mu w_\nu}{w^2}\right)\,.
\end{split}
\end{equation}

\begin{figure}
\begin{center}
\epsfig{file=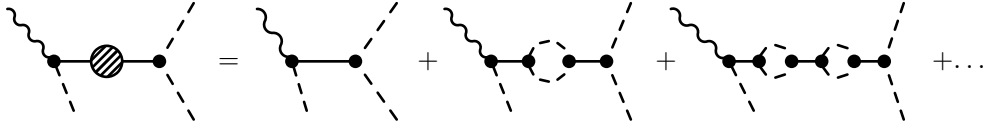, width=13cm}
\caption{Sum of diagrams, which build up the full $\rho$ propagator.}\label{fullrho1}
\end{center}
\end{figure}

\begin{figure}
\begin{center}
\epsfig{file=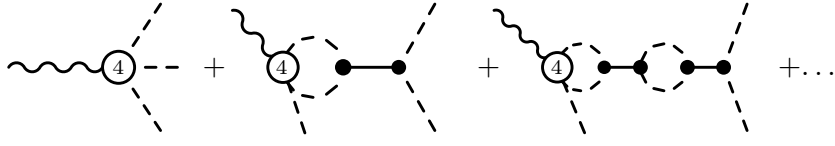, width=11cm}
\caption{Sum of higher-order diagrams, contributing to the $\tau$ decay.}\label{fullrho2}
\end{center}
\end{figure}

\begin{figure}
\begin{center}
\epsfig{file=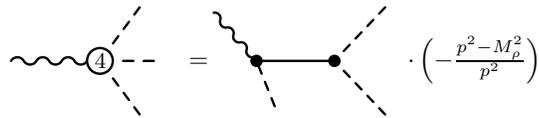, width=7cm}
\caption{Relation between higher-order contact term and lowest-order resonance diagram.}\label{fullrho3}
\end{center}
\end{figure}

\vspace{0.3cm}
\begin{center}
\textbf{Choice of interpolating fields}
\end{center}
\vspace{0.1cm}
The matrix element above was calculated by using vector fields as interpolating fields for the vector mesons. In order to improve the high-energy behaviour we include additional higher-order interactions, which account for the difference in using vector fields or tensor fields \cite{vecrep}. The additional terms are contained in Eq.(\ref{ho}), which leads to the following contribution 
\begin{figure}
\begin{center}
\begin{tabular}{cc}
\includegraphics[width=4.5cm]{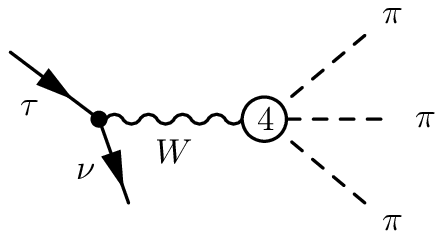} & \includegraphics[width=5cm]{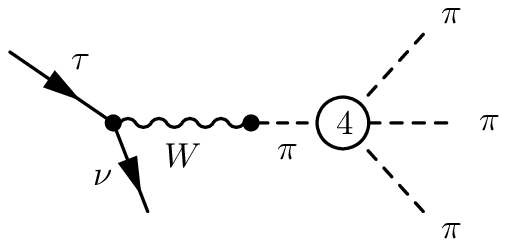} 
\end{tabular}
\caption{Higher-order contribution to the decay $\tau^- \rightarrow 2\pi^0\pi^-\nu$. The blob with label '4' denotes contact interactions emerging from Eq.(\ref{ho}). In a chiral counting these contact interactions are of fourth order.}\label{diags3}
\end{center}
\end{figure}
\begin{equation}
i\mathcal M_{ho}= C S_\mu \left(g^{\mu\nu}-\frac{w^\mu w^\nu}{M_W^2}\right)W_\nu^{3\pi ho} \,,
\end{equation}
with
\begin{equation}
\begin{split}
W^\mu_{3\pi ho} &= -\left(g^{\mu\nu} - \frac{w^\mu w^\nu}{s}\right)\biggl(\frac{gV_{ud}g_V}{ F_0^3}(f_V-2g_V)(q_{2\nu}(m_{12}\cdot q_3)-m_{12\nu}(q_2\cdot q_3)) \\
& \phantom{-\left(g^{\mu\nu} - \frac{w^\mu w^\nu}{s}\right)}+ \frac{gV_{ud}g_Vf_V}{F_0^3}(m_{12}^2 q_{\nu 2} - \frac{m_{12}^2}{2}m_{12\nu})\biggr)\\
& - \frac{2gV_{ud}g_V^2}{ F_0^3}\frac{m_\pi^2}{s(s-m_\pi^2)}w^\mu((w\cdot q_2)(m_{12}\cdot q_3)-(q_2\cdot q_3)(w\cdot m_{12})) \quad + \quad (q_1 \leftrightarrow q_3)\,,
\end{split}
\end{equation}
which is shown diagrammatically in Fig. \ref{diags3}. We will show that the net effect of adding this contribution (in a resummed way) is the replacement
\begin{equation}\label{replace}
\frac{p^2}{p^2-M_\rho^2 -\Pi} \rightarrow \frac{M_\rho^2}{p^2-M_\rho^2 -\Pi}
\end{equation}
in $W_{vec}^\mu$ (Eq.(\ref{thews3})). At a first look, it seems like we actually have to replace $p^2\rightarrow M^2_\rho + \Pi$. In order to show the validity of Eq.(\ref{replace}), we look at the sum of diagrams building up the full $\rho$ propagator, which are shown in Fig. \ref{fullrho1}.
So far we have just included additionally the two diagrams, which can be seen in Fig. \ref{diags3}. But looking at the sum in Fig. \ref{fullrho1}, it would be reasonable to also include the sum of diagrams in Fig. \ref{fullrho2}, which we call $W^\mu_{sum}$.
The central relation in this problem is that the higher-order contact term is proportional to the lowest-order diagram including the $\rho$, which is stated more precise in Fig. \ref{fullrho3}.
This relation guarantees that one can split off the $\rho\pi\pi$ vertex in the higher-order contact term in the same way as for the resonance diagram. Thus, using the relation in Fig. \ref{fullrho3}, both sums together yield
\begin{equation}
W^\mu_{vec} + W^\mu_{sum} = W_{vec}^\mu\left(1-\frac{p^2-M_\rho^2}{p^2}\right) = W_{vec}^\mu \frac{M_\rho^2}{p^2}\,,
\end{equation}
which leads exactly to the replacement advocated in Eq.(\ref{replace}).\\
We omitted the diagrams with the $\pi$ intermediate state in the discussion (right diagram Fig. \ref{diags3}), since the arguments follow exactly the same lines.

\subsection{Calculation of $\tau$ decay with explicit $a_1$}\label{a1inc}

\begin{figure}
\begin{center}
\begin{tabular}{cc}
\includegraphics[width=4.5cm]{taudecpic1} & \includegraphics[width=5cm]{taudecpic2} \\
(a) & (b)\\
\includegraphics[width=5cm]{taudecpic3} & \includegraphics[width=5.5cm]{taudecpic4} \\
(c) & (d) \\
\includegraphics[width=5.5cm]{taudecpic5} & \includegraphics[width=6cm]{taudecpic6} \\
(e) & (f) \\
\includegraphics[width=6cm]{taudecpic7} & \includegraphics[width=5.5cm]{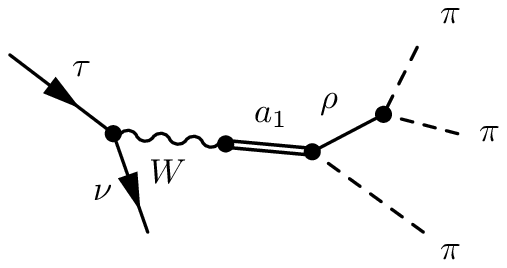} \\
(g) & (h) 
\end{tabular}
\caption{Relevant diagrams for the decay $\tau^- \rightarrow 2\pi^0\pi^-\nu$ including the explicit $a_1$.}\label{diags2}
\end{center}
\end{figure}
Next we include the $a_1$ explicitly. We introduce it as a bare resonance and generate the width by summing up the self-energy contributions from the decay of the $a_1$ in pseudoscalar and vector meson. This is automatically achieved by using the Bethe-Salpeter equation. As already mentioned, we also consider the WT term. Altogether, this leads to the diagrams shown in Fig. \ref{diags2}. In comparison to Fig. \ref{diags1}, there are two additional diagrams, where the $W$ merges into the $a_1$. Furthermore the blobs indicating the resummation are not the same as in the calculation before, since we add the $a_1$ interaction to the kernel. The additional process considered in the kernel is shown in Fig. \ref{kerna1}, and it leads to the following expression
\begin{equation}
\begin{split}
K^{\mu\nu}_{a_1}&= C_{WT}\frac{16}{F_0^2}\frac{1}{s-M_{a_1}^2}\left(g_{\alpha\beta}-\frac{w_\alpha w_\beta}{M_{a_1}^2}\right)\\
&(c_1(q^\nu p^\alpha - p\cdot q g^{\alpha\nu}) + c_2(w^\nu q^\alpha -q\cdot w g^{\alpha\nu}))(c_1(\overline q^\mu \overline p^\beta - \overline p\cdot \overline q g^{\beta\mu}) + c_2(w^\mu \overline q^\beta - \overline q\cdot w g^{\beta\mu}))\,,
\end{split}
\end{equation}
where $M_{a_1}$ is the mass of the $a_1$, which is considered to be a free parameter. The expansion into projectors is given by
\begin{align}
V_{ab11}^{1^+} &= -\frac{2}{3}F_{1ab} + g_{ab} \label{firsteq} \\
V_{ab21}^{1^+} &= -\frac{2}{3\sqrt 2 M_{Va}} (F_{1ab} \omega_a - F_{3ab} \sqrt s p_a^2) + \frac{\omega_a}{\sqrt 2 M_{Va}} g_{ab} \\
V_{ab12}^{1^+} &= \frac{2}{3\sqrt 2 M_{Vb}}(-\omega_b F_{1ab} + F_{4ab} p_b^2 \sqrt s) + \frac{\omega_b}{\sqrt 2 M_{Vb}} g_{ab} \\
V_{ab22}^{1^+} &= \frac{1}{3M_{Va} M_{Vb}}\Bigl(-\omega_a\omega_b F_{1ab} - s p_b^2 p_a^2 \frac{16 C_{WTab}}{F_0^2}\frac{1}{s-M_{a1}^2} + F_{3ab} \omega_b p_a^2 \sqrt s + F_{4ab}\omega_a p_b^2\sqrt s\Bigr) + \frac{\omega_a\omega_b}{2M_{Va}M_{Vb}} g_{ab}\,.
\end{align}
with
\begin{align}
F_1 &= \frac{16 C_{WT}}{F_0^2}\frac{1}{s-M_{a_1}^2}(c_1^2(\overline p\cdot \overline q)(p\cdot q) + c_1 c_2(\overline q\cdot w)(p\cdot q) + c_1c_2(\overline p \cdot\overline q)(q\cdot w) + c_2^2(\overline q\cdot  w)(q\cdot w))\\
F_3 &= \frac{16 C_{WT}}{F_0^2}\frac{1}{s-M_{a_1}^2}((c_1^2 - c_1 c_2) p\cdot q + (c_1 c_2 - c_2^2) q\cdot w)\\
F_4 &= \frac{16 C_{WT}}{F_0^2}\frac{1}{s-M_{a_1}^2}((c_1^2 - c_1 c_2) \overline p\cdot \overline q + (c_1 c_2 - c_2^2) \overline q\cdot w))\label{lasteq}\,.
\end{align}
We recall that the coefficient $C_{WT}$ is a matrix due to the coupled-channel structure of the problem. The indices $a,b$, which were attached to the functions $F_i$ in Eq.(\ref{firsteq}-\ref{lasteq}) correspond to the respective channels. We also note that the $F_i$ are the coefficients defined in Eq.(\ref{lor}). The whole matrix element can be written similar to Eq.(\ref{decomp})
\begin{equation}
i\mathcal M^{a1}= C S_\mu \left(g^{\mu\nu}-\frac{w^\mu w^\nu}{M_W^2}\right)W_\nu \,,
\end{equation}
but this time with
\begin{equation}
W_\nu = W_\nu^{\prime\pi\rho} + W_\nu^{\prime KK^\ast} + W_\nu^{3\pi} + W^{vec}_\nu + W_\nu^{tree} + W_\nu^{a1\pi\rho} + W_\nu^{a1KK^\ast}\,.
\end{equation}
The last two terms correspond to the diagram Fig. \ref{diags2}g, $W_\nu^{tree}$ to the process in Fig. \ref{diags2}h and $W_\nu^{\prime V\phi}$ differ from $W_\nu^{V\phi}$ because of the different kernel. The sum of the modified contributions leads to
\begin{equation}\label{wmua1}
\begin{split}
W_\mu^{Ta1} & \equiv W_\mu^{\prime\pi\rho} + W_\mu^{\prime KK^\ast} + W_\mu^{tree} + W_\mu^{a1\pi\rho} + W_\mu^{a1KK^\ast} \\
& = \left(g_{\mu\alpha}-\frac{w_\mu w_\alpha}{w^2}\right) \frac{m_{12}^2}{m_{12}^2-M_\rho^2 -\Pi}(q_1-q_2)_\delta (A_1 L_1^{\delta\alpha} + A_2 L_3^{\delta\alpha})\quad + \quad (q_1 \leftrightarrow q_3)\,.
\end{split}
\end{equation}
The tensors $L_1^{\mu\nu}$ and $L_3^{\mu\nu}$ are given by
\begin{equation}
L_1^{\mu\nu}=g^{\mu\nu}-\frac{w^\mu w^\nu}{s}, \quad L_3^{\mu\nu}=w^\mu \overline q^\nu - w^\mu w^\nu\frac{\overline q\cdot w}{s}.
\end{equation}
The coefficients $A_i$, which incorporate the nontrivial part, are given in Appendix \ref{secdetails}. The rest of the calculation follows the same lines as in the calculation before. We only have to substitute $W^\mu_{T}$ with $W^\mu_{Ta1}$ in the calculation of $W_1$. $W_2$ does not change, because neither $W^\mu_{T}$ nor $W^\mu_{Ta1}$ contribute to $W_2$ and we leave it out again.\\
At the beginning of this section we used that the $a_1$ vertex, which results from the Lagrangian Eq.(\ref{axlag}), is given by
\begin{equation}\label{a1vert}
\Gamma_{a1}^{\mu\nu}=-\frac{2 \sqrt 2 c_{\phi V}}{F_0}c_1(q^\nu p^\mu - p\cdot q g^{\mu\nu}) -\frac{2 \sqrt 2 c_{\phi V}}{F_0}c_2(w^\nu q^\mu - w\cdot q g^{\mu\nu})\,.
\end{equation}
Using $\epsilon_\nu(p)p^\nu=0$, $s=M_{a_1}^2$, $M_{a_1}^2=2M_\rho^2$ and $c_1=2c_2$ the vertex can be cast into a different form
\begin{equation}\label{oseta1vert}
\Gamma_{a1}^{\mu\nu} =\frac{2\sqrt 2 c_{\phi V}}{F_0}c_2(2q^\nu p^\mu - (2p\cdot q+w\cdot q) g^{\mu\nu} + w^\nu q^\mu) =\frac{2\sqrt 2 c_{\phi V}}{F_0}c_2(w^\nu p^\mu - w\cdot p g^{\mu\nu})\,,
\end{equation}
which is the actual vertex, which is used in \cite{oseta1}. We see that only with the simplifications above, the vertex of \cite{oseta1} is the same as the one we use. These simplifications, however, basically mean to put certain momenta onshell and apply Weinberg's relation between the $\rho$ and $a_1$ mass \cite{weinberg2}, as well as a relation between $c_1$ and $c_2$, which are anyway free parameters. We will discuss the influence of the difference between Eq.(\ref{a1vert}) and Eq.(\ref{oseta1vert}) in Section \ref{seca1res}.

\subsection{Calculation of $\tau$ decay including higher-order terms}\label{sechig}

This time we again assume that the $a_1$ is generated dynamically. In addition to the WT term, we consider higher-order corrections to the kernel of the Bethe-Salpeter equation. One part of the higher-order correction, as the WT term itself, is contained in the kinetic part of the Lagrangian and it leads to the new contribution
\begin{equation}\label{kincon}
K_1^{\mu\nu} = \frac{C_{WT}}{2 F_0^2}(q^\nu \overline q^\mu - q^\mu \overline q^\nu)\,.
\end{equation}
Next we write down all terms with two pion momenta, which one can construct, taking care of parity, C-invariance, hermiticity and of course chiral symmetry
\begin{equation}\label{holag}
\begin{split}
\mathcal L_{ho} &= \lambda_1^\prime \text{Tr}[V_\mu V^\mu u_\nu u^\nu] + \lambda_2^\prime \text{Tr}[V_\mu u_\nu V^\mu u^\nu] + \lambda_3^\prime \text{Tr}[V_\mu V_\nu u^\mu u^\nu] + \lambda_4^\prime \text{Tr}[V_\mu V^\nu u_\nu u^\mu] + \lambda_5^\prime \text{Tr}[V^\mu u_\mu V_\nu u^\nu + V_\mu u^\nu V_\nu u^\mu] \\
& + \lambda_6^\prime \text{Tr}[V_{\mu\nu} V^{\nu\alpha} u_\alpha u^\mu] + \lambda_7^\prime \text{Tr}[V_{\mu\nu} V^{\nu\alpha} u^\mu u_\alpha] + \lambda_8^\prime \text{Tr}[V_{\mu\nu} u_\alpha V^{\nu\alpha} u^\mu + V_{\mu\nu} u^\mu V^{\nu\alpha} u_\alpha] + \lambda_9^\prime \text{Tr}[V_\mu u^\nu]\text{Tr}[V^\mu u_\nu]\\
& + \lambda_{10}^\prime \text{Tr}[V_\mu u^\mu]\text{Tr}[V_\nu u^\nu] + \lambda_{11}^\prime \text{Tr}[V_\mu u_\nu]\text{Tr}[V^\nu u^\mu] + \lambda_{12}^\prime \text{Tr}[V_{\mu\alpha} u^\mu]\text{Tr}[V_\nu^\alpha u^\nu] + \lambda_{13}^\prime \text{Tr}[V_{\mu\alpha} u_\nu]\text{Tr}[V^{\nu\alpha} u^\mu]\,.
\end{split}
\end{equation} 
This together with Eq.(\ref{kincon}) leads to the following kernel in addition to the WT term
\begin{equation}
\begin{split}
K^{\mu\nu}_{ho} & = \frac{4 C_{WT}}{F_0^2}\left((\lambda_1^\prime - 2\lambda^\prime_2)(\overline q\cdot q)g^{\mu\nu} + (\lambda_3^\prime-2\lambda_5^\prime-\frac{1}{2}) \overline q^\nu q^\mu + (\lambda_4^\prime-2\lambda_5^\prime+\frac{1}{2}) \overline q^\mu q^\nu - (\lambda_6^\prime + \lambda_7^\prime - 2\lambda_8^\prime)(w\cdot q)(w\cdot\overline q)g^{\mu\nu}\right)\\
& -\frac{8\mathds{1}_2}{F_0^2}((\lambda_9^\prime (q\cdot\overline q) + (\lambda_{12}^\prime + \lambda_{13}^\prime)(w\cdot q)(w\cdot\overline q))g^{\mu\nu} + \lambda_{10}^\prime q^\mu \overline q^\nu + \lambda_{11}^\prime \overline q^\mu q^\nu) \,.
\end{split}
\end{equation}
We see that there are only eight independent variables contributing to the process. Dropping terms, which will not contribute to $J^P=1^+$, we are down to six independent variables, which we call $\lambda_1,\lambda_2,\dots\lambda_6$. Thus, the kernel can be written as
\begin{equation}\label{hokern}
\begin{split}
K_{ho}^{\mu\nu} &= \frac{C_{WT}}{F_0^2}\Bigl(\bigl(\lambda_1(q\cdot \overline q) + \lambda_2(w\cdot q)(w\cdot \overline q)\bigr)g^{\mu\nu} + \lambda_3 q^\mu \overline q^\nu \Bigr)\\
& - \frac{\mathds{1}_2}{F_0^2}\Bigl(\bigl(\lambda_4(q\cdot\overline q) + \lambda_5(w\cdot q)(w\cdot\overline q)\bigr)g^{\mu\nu} + \lambda_6 q^\mu \overline q^\nu \Bigr) \,.
\end{split}
\end{equation}
Together with the WT term the expansion of the kernel reads
\begin{align}
V_{ab11}^{1^+} &= -\frac{2}{3}F_{1ab} + g_{ab} \\
V_{ab21}^{1^+} &= -\frac{2}{3\sqrt 2 M_{Va}} (F_{1ab} \omega_a - F_{3ab} \sqrt s p_a^2) + \frac{\omega_a}{\sqrt 2 M_{Va}} g_{ab} \\
V_{ab12}^{1^+} &= \frac{2}{3\sqrt 2 M_{Vb}}(-\omega_b F_{1ab} + F_{4ab} p_b^2 \sqrt s) + \frac{\omega_b}{\sqrt 2 M_{Vb}} g_{ab} \\
V_{ab22}^{1^+} &= \frac{1}{3M_{Va} M_{Vb}}\biggl(-\omega_a\omega_b F_{1ab} - (\lambda_1 C_{WTab}-\lambda_4 \delta_{ab}) \frac{1}{F_0^2} p_b^2 p_a^2 + F_{3ab} \omega_b p_a^2 \sqrt s + F_{4ab}\omega_a p_b^2\sqrt s\biggr) + \frac{\omega_a\omega_b}{2M_{Va}M_{Vb}} g_{ab}\,.
\end{align}
with
\begin{align}
F_1 &= \frac{C_{WT}}{F_0^2}(q_0\overline q_0)(\lambda_1 + s\lambda_2) - \frac{\mathds{1}_2}{F_0^2}(q_0\overline q_0)(\lambda_4 + s\lambda_5) \\
F_3 &= \frac{C_{WT}}{F_0^2}\frac{q\cdot w}{s}\lambda_3 - \frac{\mathds{1}_2}{F_0^2}\frac{q\cdot w}{s}\lambda_6\\
F_4 &= \frac{C_{WT}}{F_0^2}\frac{\overline q\cdot w}{s}\lambda_3 - \frac{\mathds{1}_2}{F_0^2}\frac{\overline q\cdot w}{s}\lambda_6\,.
\end{align}
The diagrams, we have to include are the same as in Fig. \ref{diags1} with a different scattering amplitude describing the final state correlations, which leads to
\begin{equation}
i\mathcal M_{\phi V}^{ho}= C S_\mu \left(g^{\mu\nu}-\frac{w^\mu w^\nu}{M_W^2}\right)W_\nu \,,
\end{equation}
with
\begin{equation}
W_\nu = W_\nu^{\prime\prime\pi\rho} + W_\nu^{\prime\prime KK^\ast} + W_\nu^{3\pi} + W^{dir}_\nu \,,
\end{equation}
where
\begin{equation}
\begin{split}
W_\mu^{\prime\prime \phi V} &= \frac{g V_{ud} g_V}{2 \sqrt 2 F_0^3}c_{\phi V} J_{\phi V}(\mu_2)\left(g_{\mu\alpha}-\frac{w_\mu w_\alpha}{w^2}\right)
\Biggl(\left(g_V \alpha_1^{\phi V} + \frac{1}{2}(f_V-2g_V)\alpha_2^{\phi V}\right) L_1^{\gamma\alpha} \\
&+ \left(g_V \alpha_3^{\phi V} + \frac{1}{2}(f_V-2g_V)\alpha_4^{\phi V}\right) L_3^{\gamma\alpha}\Biggr)\frac{m_{12}^2}{m_{12}^2-M_\rho^2 -\Pi}(q_1-q_2)_\gamma + \quad (q_1 \leftrightarrow q_3)\,,
\end{split}
\end{equation}
The coefficients $\alpha_i$ can be found in Appendix \ref{secdetails}. They are the same as in the case of including the explicit $a_1$ except that one has to replace the expansion coefficients of the scattering amplitude $M_{abij}$.

\subsection{$W$ form factor}\label{secalt}

\begin{figure}
\begin{center}
\epsfig{file=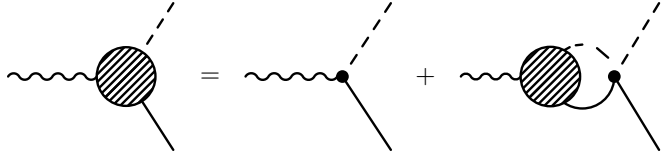, height=2cm}
\caption{Diagrammatic form of the equation to determine the form factor of the $W$ boson. The dashed lines represent pions, the solid lines the $\rho$ mesons and the wiggly line the $W$ boson. The bare vertex includes also the diagram with the intermediate pion (see Eq.(\ref{eqbare})).}\label{wformpic}
\end{center}
\end{figure}

\begin{figure}
\begin{center}
\epsfig{file=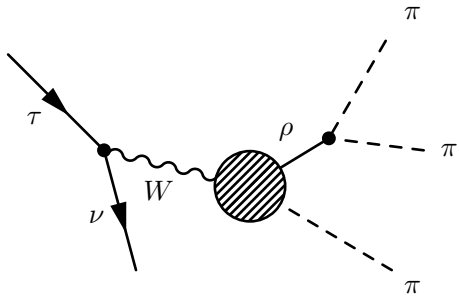, width=6cm}
\caption{Diagram describing the $\tau$ decay in the simplified version.}\label{wformdiag}
\end{center}
\end{figure}

Instead of first calculating the scattering amplitude, one could introduce the $W$ form factor to determine the decay. Leaving out some details and only considering the WT term, it is possible to work out the decay width in a few lines. It is instructive to look at this simple calculation, since here the intermediate steps are not clouded by lengthy algebra and the core of the calculation is better visible.\\
'Leaving out details' means
\begin{itemize}
\item neglect longitudinal part of the hadronic tensor proportional to $m_\pi^2$
\item $f_V-2g_V=0$
\item neglect lowest-order CHPT diagrams (direct three-pion decays)
\item only $\pi\rho$ channel, no coupled channels
\end{itemize}
The first simplification actually has no visible influence on the result. The second simplification is numerically almost fulfilled, i.e. 
\begin{equation}\label{fvest}
\frac{f_V-2g_V}{f_V} \approx 0.1\quad \rightarrow \quad f_V >> (f_V-2g_V)
\end{equation}
The third approximation will only influence the very low-energy region of the decay. The only serious simplification is the last one, which we will later also see to have a minor influence on the results. Therefore, we can expect this slimmed down version to be pretty close to the full calculation.\\
The equation, determining the $W$ form factor, is
\begin{equation}\label{wform}
V^{\mu\nu}(q,w) = V^{\mu\nu}_0(q,w) + \int\frac{d^4 l}{(2\pi)^4}V^{\mu\alpha}(l,w)G_{\alpha\beta}(l,w)K^{\beta\nu}(l,q,w)\,,
\end{equation}
which can be seen in pictorial form in Fig. \ref{wformpic}.
With the simplifications above, $V_0^{\mu\nu}$ is given by
\begin{equation}\label{eqbare}
V_0^{\mu\nu} =\frac{-i g V_{ud} f_V p^2}{2 F_0}\left(g^{\mu\alpha} - \frac{w^\mu w^\alpha}{s}\right)(g_\alpha^\nu - \frac{p_\alpha p^\nu}{p^2})\,,
\end{equation}
where $p$ is the momentum of the vector meson. We drop the term proportional to $p^\nu$, since it will not contribute due to the form of the $\rho\pi\pi$ vertex ($\sim (q_1-q_2)^\mu$) and the renormalisation scheme, in which tadpoles are dropped. Thus, we get
\begin{equation}
V_0^{\mu\nu} = \frac{-i g V_{ud} f_V M_\rho^2}{2 F_0} \left(g^{\mu\nu}-\frac{w^\mu w^\nu}{s}\right) \equiv V_0 L_1^{\mu\nu}
\end{equation}
with $L_1^{\mu\nu}$ defined in Eq.(\ref{thels}). The kernel $K^{\mu\nu}$ is already known to be (see Eq.(\ref{wtkern}))
\begin{equation}
K^{\mu\nu} = K_0 L_1^{\mu\nu}
\end{equation}
with 
\begin{equation}
K_0 =  -\frac{1}{4F_0^2}\left(3s - (2 m_\pi^2 + 2M_\rho^2) -\frac{1}{s}(M^2_\rho - m_\pi^2)^2\right)\,.
\end{equation}
Looking at Eq.(\ref{wform}) and the form of the kernel, i.e. that it does not depend on $q$, we can write down a reasonable ansatz for $V_{\mu\nu}$
\begin{equation}
V^{\mu\nu} = V(s) L_1^{\mu\nu}\,.
\end{equation}
Plugging in this ansatz in Eq.(\ref{wform}) we get
\begin{equation}
V L_1^{\mu\nu} = V_0 L_1^{\mu\nu} + K_0 V I_{\pi\rho}\left(\frac{2}{3} + \frac{1}{12M_\rho^2 s}(m_\pi^2-M_\rho^2-s)^2\right) L_1^{\mu\nu}\,,
\end{equation}
and we can easily read off $V$ to be
\begin{equation}
V=\frac{V_0}{1 - K_0\left(\frac{2}{3}+\frac{1}{12M_\rho^2 s}(m_\pi^2-M_\rho^2-s)^2\right)I_{\pi\rho}}\,.
\end{equation}
The result is rendered finite by substituting $I_{\pi\rho} \rightarrow J_{\pi\rho}(\mu_1)$. The above calculation seems to employ only one subtraction point $\mu_1$. This is in contrast to our derivation in Section \ref{secwwt}, where we argued that two different subtraction points can appear, namely one to renormalise the Bethe-Salpeter equation and one for the entrance loop from the $W$ boson into the rescattering process. We will show in the following how the second subtraction point can also be recovered in the present calculation.\\
Omitting the Lorentz structure for the moment, the full $W$ decay vertex can be written as
\begin{equation}
V = V_0 + V_0 G T\,.
\end{equation}
In the solution of the Bethe-Salpeter equation for the scattering matrix $T$, we will use $G^\prime$ in the following, in order to indicate a possibly different subtraction point. Using $T=(1-KG^\prime)^{-1}K$ the form factor $V$ can be written as
\begin{align}
V &= V_0 + V_0 G (1-KG^\prime)^{-1}K = V_0 (1-G^\prime K)(1-G^\prime K)^{-1} + V_0 G K(1-G^\prime K)^{-1}\nonumber \\
& = V_0(1-G^\prime K + GK)(1-G^\prime K)^{-1}\,,
\end{align}
which corresponds to an equation of the form shown in Fig. \ref{wformpic}, provided one takes the bare $W$ form factor as 
\begin{equation}\label{vprime}
V_0^\prime =  V_0(1-G^\prime K + GK)\,.
\end{equation}
We see that in the vertex we can effectively include a change in the subtraction point of the first loop relative to the subtraction point of the Bethe-Salpeter equation. We note that the change in the subtraction point is at least one order higher in a chiral counting, since the kernel $K$ is already $\mathcal O(q)$. We recall that Eq.(\ref{lagvec}) contains only the lowest-order $W\rightarrow \phi V$ vertex. Using different renormalisation points for the loops $G$ and $G^\prime$ gives us the possibility to account for modifications of this lowest-order expression. \\
For the calculation of the whole decay, we only use one diagram, which is shown Fig. \ref{wformdiag}. 
Thus, we get
\begin{equation}
i \mathcal M = C S^\mu \left(g_{\mu\nu}-\frac{w_\mu w_\nu}{M_W^2}\right)W^\nu_{form} 
\end{equation}
with
\begin{equation}\label{wmuform}
W^\mu_{form} = -\frac{i V g_V}{F_0^2} \frac{m_{12}^2}{m_{12}^2-M_\rho^2 - \Pi}L_1^{\mu\alpha}(q_{1\alpha}-q_{2\alpha}) \quad + \quad  (q_1 \leftrightarrow q_3)\,.
\end{equation}
We do not show the results of the simplified calculations explicitly, since one can anticipate the outcome by looking at the discussions in Section \ref{secres}. In particular, in Fig. \ref{wokk} we will see that neglecting the strangeness channel does not have a big effect. The other simplifications have already been estimated above to be less important.

\section{Results}\label{secres}

\subsection{Calculation without $a_1$}\label{secwoa1}

First we want to investigate the spectral function for the decay $\tau^- \rightarrow 2\pi^0 \pi^- \nu_\tau$ calculated by iterating the WT term in order to dynamically generate the $a_1$. We will discuss the influence of different aspects of the calculation on the results in detail and determine the values of the subtraction points.

\begin{itemize}
\item Influence of interpolating fields and spectral distribution
\end{itemize}
We discussed the different possibilities to describe vector particles, namely in terms of vector fields and in terms of antisymmetric tensor fields. We introduced higher-order corrections in order to account for the difference stemming from the choice of fields. For the present calculation we note that using the antisymmetric tensor fields, leads to the appearance of less derivatives, and therefore we expect a better high-energy behaviour. Instead of explicitly using the antisymmetric tensor fields, we use the vector representation but also include the higher-order contact terms given in Eq.(\ref{ho}).
\begin{figure}
\begin{center}
\includegraphics[width=9cm]{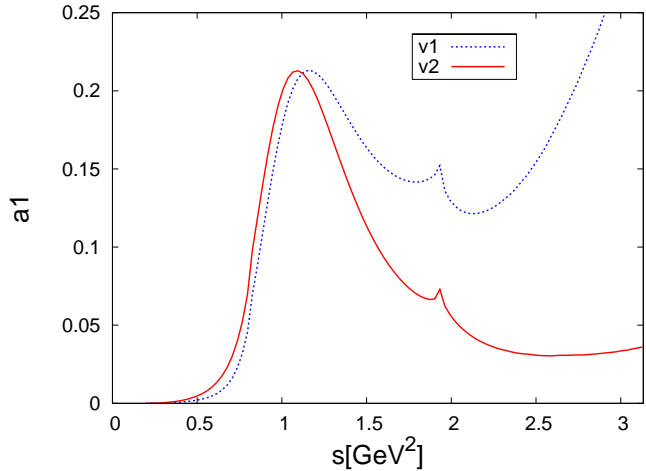}
\caption{Spectral function for the decay $\tau^-\rightarrow 2\pi^0 \pi^-\nu_\tau$ calculated with different choices for the interpolating fields. v1 only uses the vector field Lagrangian Eq.(\ref{lagvec}), whereas v2 additionally includes the contact terms from Eq.(\ref{ho}).}\label{vec}
\end{center}
\end{figure}
Fig. \ref{vec} shows the spectral function calculated with vector fields (v1) and with vector fields including the higher-order terms (v2). One clearly sees the better high-energy behaviour for the case of v2. In Fig. \ref{vec} we used dimensional regularisation with $\mu_1=\mu_2=M_\rho^2$, which is the value from \cite{lutz2}. We will discuss the influence of the subtraction points in detail below. The kink which can be seen at about $1.9\,$GeV$^2$ results from the threshold of the $KK^\ast$ channel. Using spectral distributions for the vector mesons, taken from \cite{spec}, smoothes the curve, which can be seen in Fig. \ref{rencomp2}.
\begin{figure}
\begin{center}
\includegraphics[width=9cm]{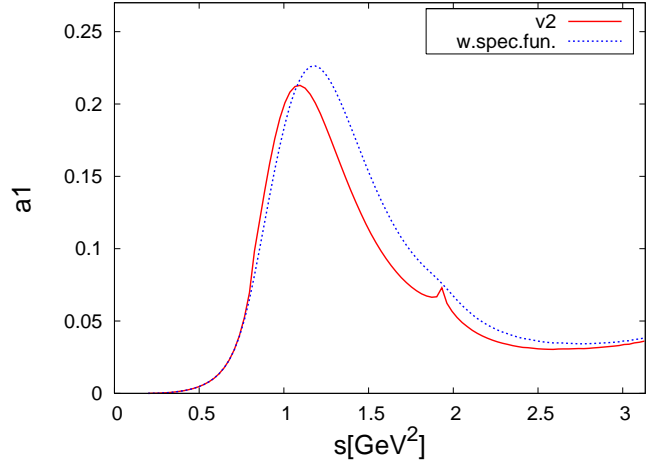}
\caption{Spectral function for the decay $\tau^-\rightarrow 2\pi^0 \pi^-\nu_\tau$ calculated with and without including the width of the vector mesons in the loop integral. The curve labelled v2 is the same as in Fig. \ref{vec}.}\label{rencomp2}
\end{center}
\end{figure}
The curve also gets a little broader and moves to the right, but the overall structure is unchanged. If we do not state otherwise, the following calculations will always contain the higher-order corrections (i.e. v2) and the spectral function for the vector mesons in the loop.

\begin{figure}
\begin{center}
\includegraphics[width=9cm]{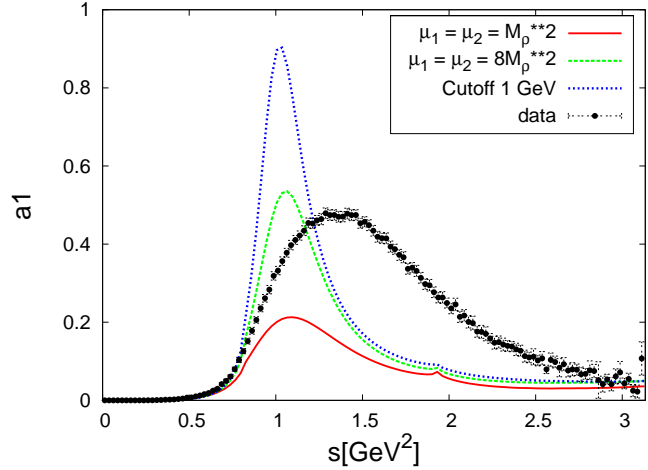}
\caption{Spectral function for the decay $\tau^-\rightarrow 2\pi^0 \pi^-\nu_\tau$ calculated with different renormalisation schemes in comparison to data from \cite{aleph1}.}\label{rencomp1}
\end{center}
\end{figure}

\begin{figure}
\begin{center}
\begin{tabular}{cc}
\includegraphics[width=7cm]{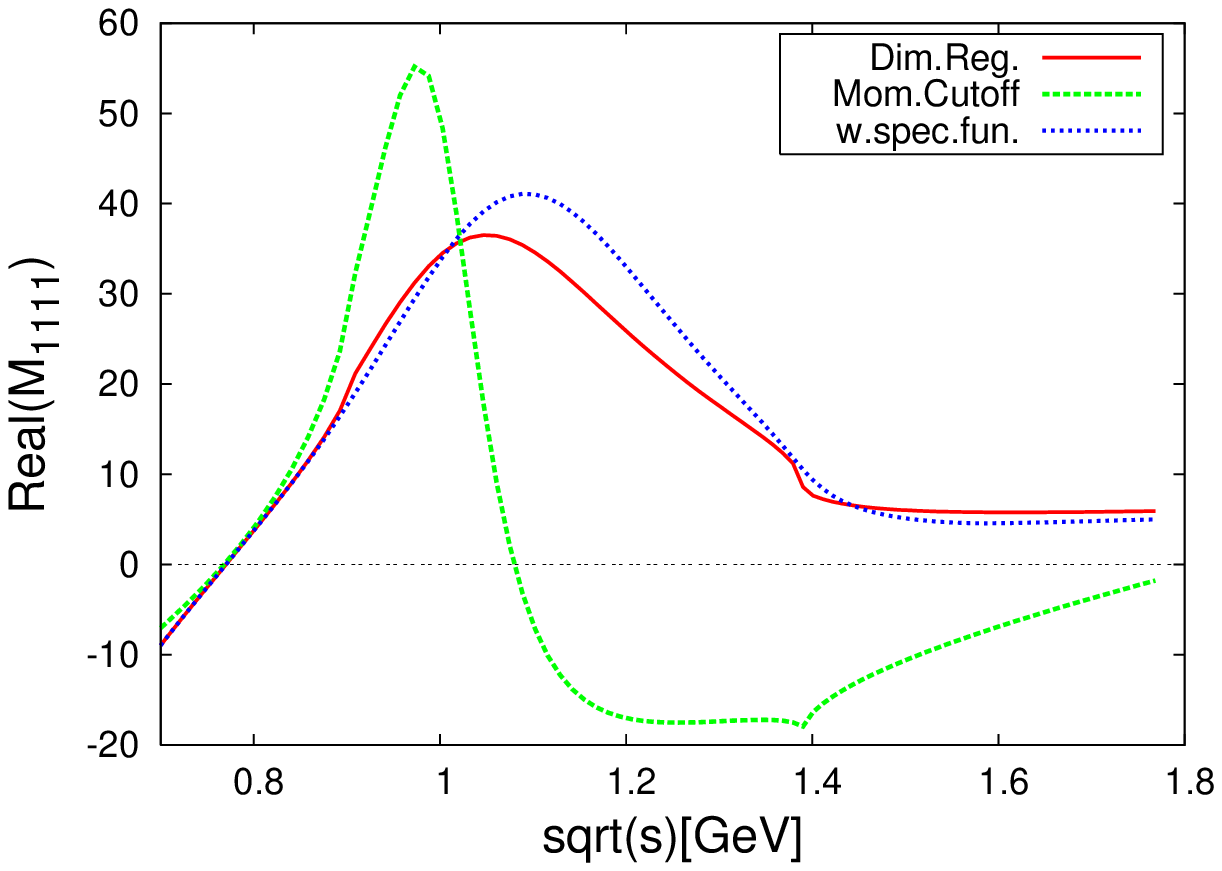} & \includegraphics[width=7cm]{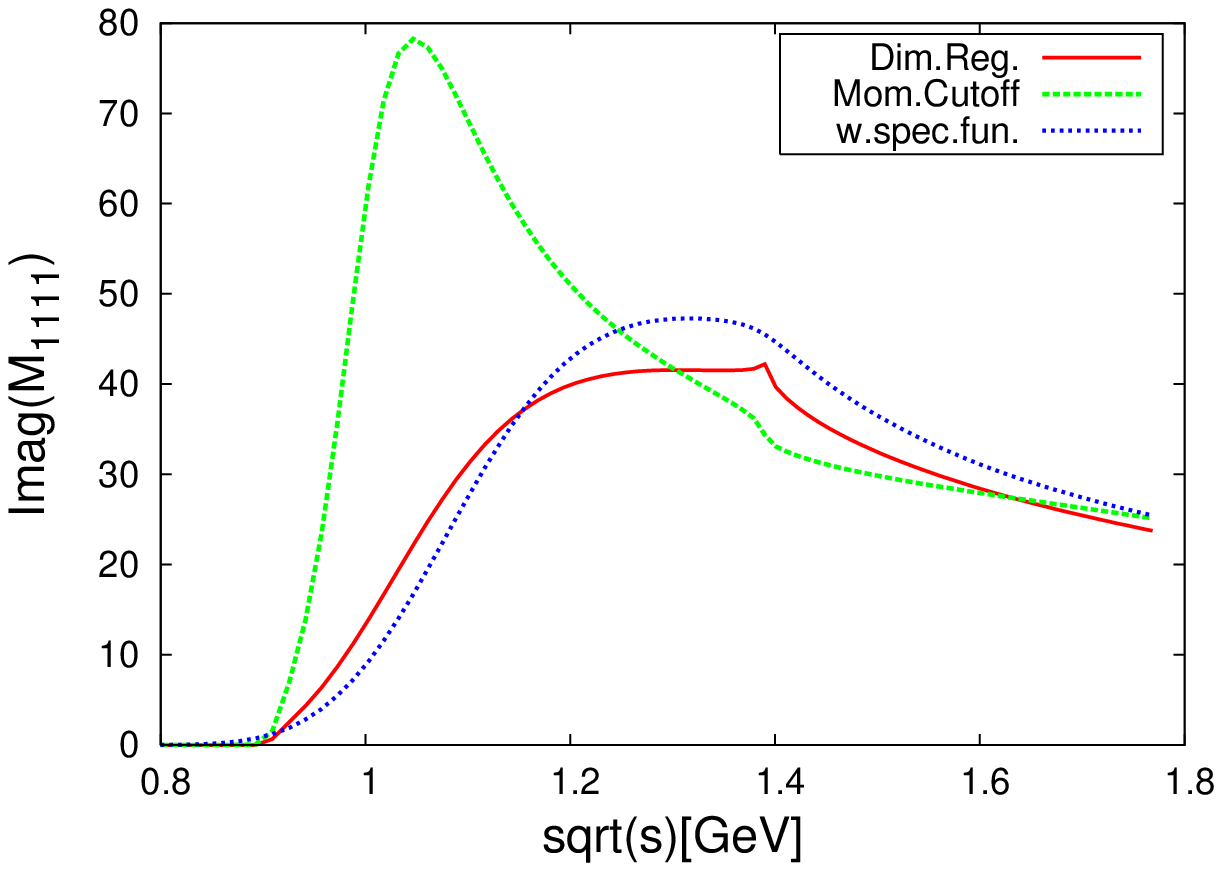}
\end{tabular}
\caption{Real (left) and imaginary (right) part of the scattering amplitude for $\pi\rho$ scattering. The curves correspond to the highest curve in Fig. \ref{rencomp1} and to the two curves shown in Fig. \ref{rencomp2}, where the curve labelled 'v2' in Fig. \ref{rencomp2} corresponds to the curve labelled 'Dim.Reg.' in this picture.}\label{matel}
\end{center}
\end{figure}

\begin{itemize}
\item Influence of renormalisation
\end{itemize}
In Section \ref{secdec} we encountered two subtraction points in our calculation. In the following we will investigate the influence of these two parameters. We start by setting $\mu_1=\mu_2$ and vary them simultaneously. In Fig. \ref{rencomp1}
we see the spectral function for the decay $\tau^-\rightarrow 2\pi^0 \pi^-\nu_\tau$ calculated with different renormalisation points in comparison to data from \cite{aleph1}. The lowest curve is calculated using dimensional regularisation with $\mu_1=\mu_2=M_\rho^2$, which corresponds to the value employed in \cite{lutz2}. For the curve, which turned out to be the highest, we choose a cutoff scheme with a cutoff at 1$\,$GeV in momentum, which corresponds to the choice in \cite{osetscalar}. The result in between was again calculated with dimensional regularisation, but this time with a subtraction point at $\mu_1=\mu_2=8.5M_\rho^2$. Since including the spectral function for the vector mesons in the cutoff scheme is more complicated and does not lead to new insights for this comparison, we did not include the spectral distribution for the calculations shown in Fig. \ref{rencomp1}. The different curves clearly differ in the height of the peak, but the position of the peak is not influenced. The width of the peak turns out to be too small in all prescriptions.\\
It is also instructive to look directly at the scattering amplitudes, which describe the rescattering. We note that the scattering amplitude only depends on $\mu_1$ and is independent of $\mu_2$. In Fig. \ref{matel} we see the real and imaginary part of the scattering amplitude for $\pi\rho$ scattering (corresponding to $M_{1111}$) for different renormalisation descriptions. The curves shown in Fig. \ref{matel} correspond to the highest and lowest curve in Fig. \ref{rencomp1}. In addition, we plotted the lowest curve using a spectral distribution for vector mesons in the loop, which corresponds to the result in Fig. \ref{rencomp2}.
We see that the scattering amplitude hardly shows a resonance structure by using the subtraction point at $M_\rho^2$ (full curve in Fig. \ref{matel}). The resonant structure is more pronounced for the cutoff scheme (dashed curve in Fig. \ref{matel}). Including the spectral function of the $\rho$ (dotted curve in Fig. \ref{matel}) basically smoothes the curve. The bump in the imaginary part is moved to the left in the cutoff scheme, whereas in Fig. \ref{rencomp1} one could hardly see a difference in the position of the peak. This shows that it is not so obvious to translate the structure seen in the scattering amplitude to the spectral function of the $\tau$ decay. In other words, interferences between the tree level diagrams and the rescattering diagrams (cf. Fig. \ref{diags1}) play an important role.\\
In the following we will only use dimensional regularisation to render the loops finite. With this restriction we nevertheless cover the full discussion on the renormalisation parameter since there are only marginal differences by using different schemes, provided that the parameters are properly chosen. In principle this can be seen in Fig. \ref{rencomp1}, where one can easily imagine that a further increase of the subtraction point leads to the same result as the one which was calculated with the cutoff.

\begin{figure}
\begin{center}
\includegraphics[width=9cm]{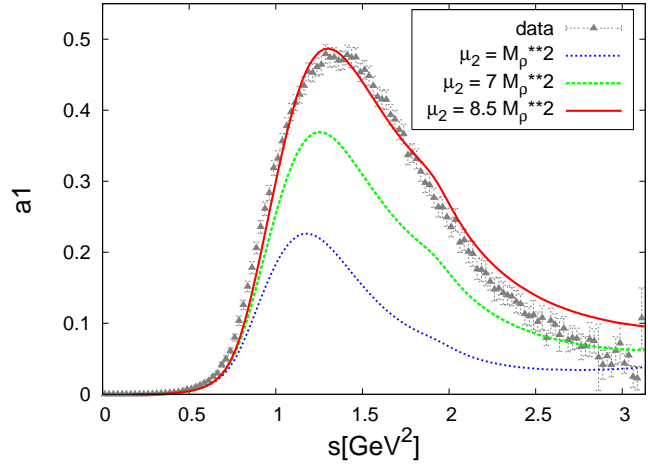}
\caption{Spectral function for the decay $\tau^-\rightarrow 2\pi^0 \pi^-\nu_\tau$, calculated by varying the subtraction point $\mu_2$ of the first loop and keeping the subtraction point in the scattering amplitude fixed at $\mu_1=M_\rho^2$, in comparison to data from \cite{aleph1}.}\label{sub2p}
\end{center}
\end{figure}

\begin{figure}
\begin{center}
\includegraphics[width=9cm]{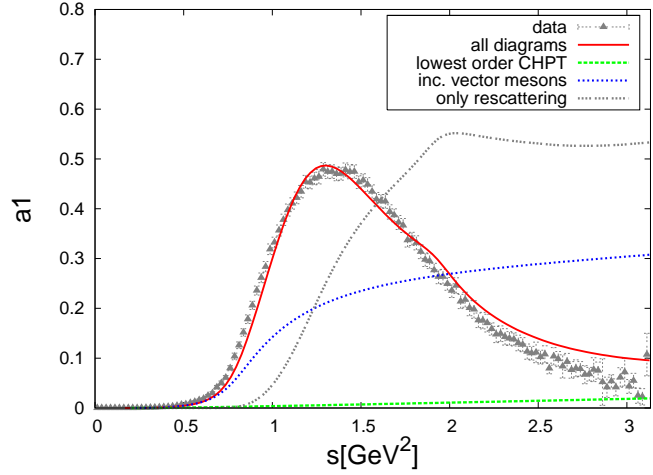}
\caption{Spectral function for $\mu_1=M_\rho^2$ and $\mu_2=8.5M_\rho^2$ (cf. Fig. \ref{sub2p}) split up into the different contributions from the diagrams in Fig. \ref{diags1} in comparison to data from \cite{aleph1}. The 'lowest order CHPT' curve corresponds to Fig. \ref{diags1}a,b, 'inc. vector mesons' to Fig. \ref{diags1}c,d and 'only rescattering' to Fig. \ref{diags1}e,f.}\label{split}
\end{center}
\end{figure}

Next we want to investigate the effect of changing $\mu_2$ while keeping $\mu_1$ fixed. We will use $\mu_1=M_\rho^2$, which in \cite{lutz2} was determined by using crossing symmetry arguments. Thus, using these arguments to fix one subtraction point, we are in principle left with only one free parameter. The results for different choices of $\mu_2$ can be seen in Fig. \ref{sub2p}.
We see that we can describe the data very well by varying only that subtraction point and keeping $\mu_1=M_\rho^2$ fixed in the scattering amplitude \cite{unsers}. We note that choosing the subtraction point at $\mu_2=8.5M_\rho^2$, which describes the data best, corresponds approximately to a cutoff of $1\,$GeV in a cutoff scheme. Obviously this value is very reasonable.\\
In Fig. \ref{split} we see the spectral function for $\mu_1=M_\rho^2$ and $\mu_2=8.5M_\rho^2$ (cf. Fig. \ref{sub2p}) split into the different contributions from the diagrams shown in Fig. \ref{diags1}.
We see that the bump is partly created by the negative interference of the rescattering diagrams and the diagrams including the vector mesons at tree level. The little bump, we see in the rescattering contribution alone appears at the wrong position and only the sum of all diagrams gives the pronounced peak, which is of course the only quantity that can be measured.

\begin{itemize}
\item Influence of coupled channels
\end{itemize}
\begin{figure}[ht]
\begin{center}
\includegraphics[width=9cm]{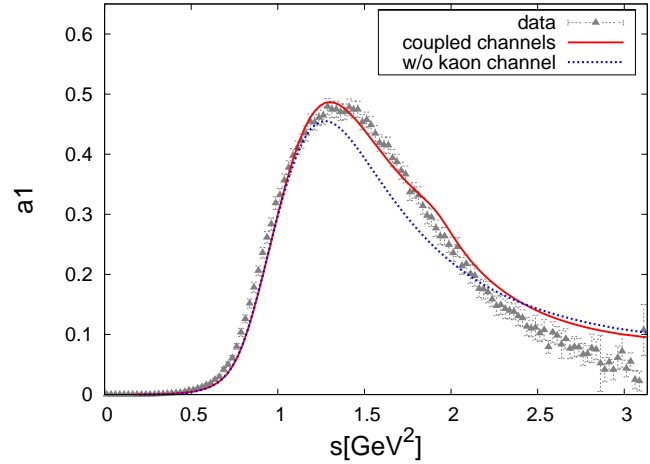}
\caption{Spectral function for the decay $\tau^-\rightarrow 2\pi^0 \pi^-\nu_\tau$ calculated with and without including the kaon channel in comparison to data from \cite{aleph1}.}\label{wokk}
\end{center}
\end{figure}
The spectral function calculated with and without including the strangeness channel is shown in Fig. \ref{wokk}, which shows that the bump also appears without the kaon channel. The height is a little less by leaving out the kaons, but that could be compensated by varying the subtraction point $\mu_2$. The rise in the data in the energy region up to about 1.1$\,$GeV$^2$ can also be described by leaving out the kaons, but the width of the peak is better described by including both channels. However, the effect is pretty small and one can safely say that the $\pi\rho$ channel plays the dominant role.

\begin{itemize}
\item Varying $g_V$ and $f_V$
\end{itemize}
\begin{figure}[ht]
\begin{center}
\includegraphics[width=9cm]{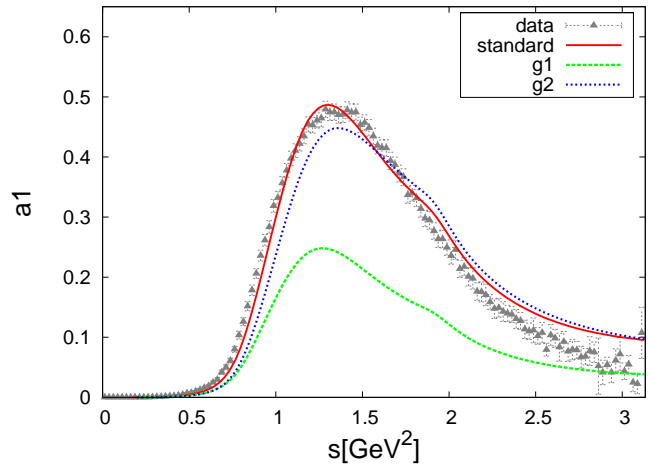}
\caption{Spectral function for the decay $\tau^-\rightarrow 2\pi^0 \pi^-\nu_\tau$ calculated with different values for $g_V$ and $f_V$ in comparison to data from \cite{aleph1}. 'g1' corresponds to the theoretically motivated values of $f_V$ and $g_V$ according to Eq.(\ref{fvgvres}) with the subtraction points $\mu_1=M_\rho^2$ and $\mu_2=8.5M_\rho^2$. 'g2' uses the same parameters as 'g1', except $\mu_2$, which is chosen to be $\mu_2=14M_\rho^2$.}\label{gv}
\end{center}
\end{figure}
So far we used the experimentally measured values for $f_V$ and $g_V$, which are given by
\begin{equation}\label{fvgvres1}
f_V = \frac{0.154\,\text{GeV}}{M_\rho}\,,\quad g_V=\frac{0.069\,\text{GeV}}{M_\rho}\,.
\end{equation}
As already noted in Section \ref{secinter}, these values slightly differ from  
\begin{equation}\label{fvgvres}
f_V = 2g_V\,,\quad g_V=\frac{F_0}{\sqrt 2 M_\rho}\,,
\end{equation}
which are the values, obtained by theoretical considerations and approximations in \cite{vecrep}. In order to see the influence of these parameters on the results, we show in Fig. \ref{gv} the spectral function calculated with different values for $g_V$ and $f_V$. 'g1' corresponds to the theoretically motivated values of $f_V$ and $g_V$ according to Eq.(\ref{fvgvres}) with the subtraction points $\mu_1=M_\rho^2$ and $\mu_2=8.5M_\rho^2$. 'g2' uses the same parameters as 'g1', except for $\mu_2$, which is chosen to be $\mu_2=14M_\rho^2$ in order to fit approximately the height of the peak (and still get roughly the shape). We see that the moderate difference in $f_V$ and $g_V$ has a sizeable impact on the results, which is due to the fact that the combination $f_V g_V$ appears quadratically in the final formulas. The difference of $f_V g_V$ between using the values of Eq.(\ref{fvgvres1}) and Eq.(\ref{fvgvres}) is about $25\%$. The change in the height of the peak can be compensated by a readjustment of the subtraction point $\mu_2$, but the spectral function in this case seems to be shifted to the right. We note that there is no other parameter, which potentially can influence the spectral function up to about $s\approx 0.7\,$GeV$^2$, as can be seen from the discussions before.\\
Except of the influence at low energies and the resulting small shift, varying $f_V$ and $g_V$ seems to have a similar effect as varying $\mu_2$. This is not too surprising, since changing $f_V$ and $g_V$ changes also the $W$ decay vertex and leaves the scattering amplitude untouched (cf. discussion in Section \ref{secalt}).

\begin{itemize}
\item Stable $\rho$
\end{itemize}
\begin{figure}[ht]
\begin{center}
\includegraphics[width=9cm]{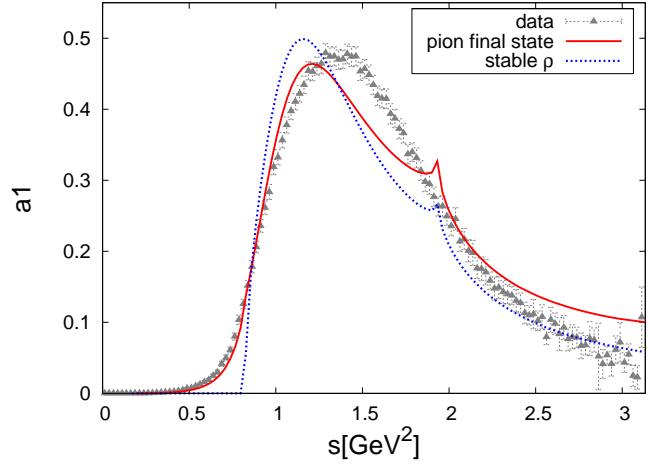}
\caption{Spectral function for the decay $\tau^-\rightarrow \pi^0 \rho^-\nu_\tau$ calculated by assuming the $\rho$ to be stable ('stable $\rho$') and spectral function for the usual three-pion final state ('pion final state') in comparison to data from \cite{aleph1}. In order to see the mere difference by assuming different final states, we do not use a spectral distribution for the vector mesons in the loop. The subtraction points are chosen according to the best choice at $\mu_1=M_\rho^2$ and $\mu_2=8.5M_\rho^2$.}\label{stable}
\end{center}
\end{figure}
In Section \ref{secwhich} we discussed which diagrams we should include in our calculation, and we assumed that the contribution from the pion final state interactions is small. In order to show that this is a reasonable assumption, we compare our previous calculations with one, where the $\rho$ is assumed to be stable. This means we look at a spectral function obtained from the final state $\pi\rho$ instead of $3\pi$. With the notation from Section \ref{secdec} and neglecting the longitudinal part proportional to $m_\pi^2$, we get
\begin{equation}
W^\mu_{stable} = -\left(g^{\mu\nu}-\frac{w^\mu w^\nu}{s}\right) b_T\frac{F_0^2}{g_V}\epsilon_\nu^\rho(p) + \left(g^{\mu\nu}-\frac{w^\mu w^\nu}{s}\right)\frac{g V_{ud} f_V}{2 F_0}(p^2 g_{\nu\alpha} - p_\alpha p_\nu)\epsilon^\alpha_\rho(p)\,,
\end{equation}
and therefore
\begin{equation}
\begin{split}
W_1 =-\frac{1}{3}\frac{p_{cm}}{2 (2\pi)^5\sqrt s} &\left(|b_T|^2 \frac{F_0^4}{g_V^2} + \frac{g^2 V_{ud}^2 f_V^2}{4F_0^2}M_\rho^4 - \frac{f_V}{g_V}F_0 g V_{ud}M_\rho^2 \Re(b_T)\right)\\
& \phantom{aa} \cdot\left(2 + \frac{1}{4sM_\rho^2}(s+M_\rho^2-m_\pi^2)^2\right)\,.
\end{split}
\end{equation}
For simplicity, we also neglected terms $\sim (f_V-2g_V)$ in the tree level diagram ($W^\mu_{vec}$). We checked explicitly, that these terms influence the results by less than $10\%$, and therefore they would only lengthen the formulas above. The negligible influence is of course expected from Eq.(\ref{fvest}). In order to see the net effect of assuming a stable $\rho$, we do not include the spectral function for the vector mesons in the loops. When the $\rho$ is stable, the threshold for the decay of the $\tau$ is of course sharper and moved to the right, which can be seen in Fig. \ref{stable}. We see that beside these differences, the structure is the same as before and there is not much room for a big contribution from pion final state interactions.

\vspace{0.5cm}

We want to summarise shortly what we saw so far. In a scenario where the $a_1$ is generated dynamically we employ two parameters $\mu_1$ and $\mu_2$. Varying both simultaneously with $\mu_1=\mu_2$, we saw that we always got a peak at the same position with varying height and a too small width. Using the value from \cite{lutz2} to fix $\mu_1$, we describe the data very well by choosing the only remaining free parameter $\mu_2$ at $8.5M_\rho^2$. We also saw that in the process under consideration the main contribution came from the $\pi\rho$ channel, while the kaon channel plays a minor role. In addition, we investigated the influence of the parameters $f_V$ and $g_V$ and found that the results are quite sensitive to these parameters. In particular, the low-energy behaviour is best described by using the values, which are directly determined from experiment. Finally we assumed the $\rho$ to be stable, which yields a qualitatively similar result. This eliminates concerns about possibly large final state interactions of the pions (cf. Fig. \ref{notinc}).

\subsection{Calculation with explicit $a_1$}\label{seca1res}

\begin{figure}
\begin{center}
\includegraphics[width=9cm]{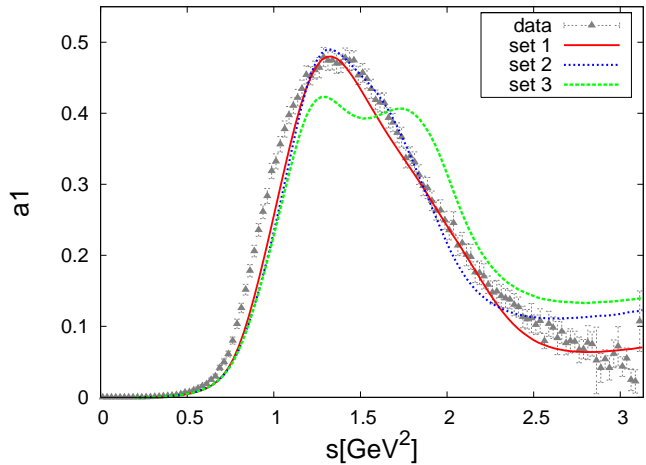}
\caption{Spectral function for the decay $\tau^-\rightarrow 2\pi^0 \pi^-\nu_\tau$ including the $a_1$ with different sets of paramaters in comparison to data from \cite{aleph1}. The parameter sets are given in Tab. \ref{a1paras}.}\label{a1}
\end{center}
\end{figure}

Now we want to look at the results of the calculation when we include the $a_1$ explicitly. A very small coupling of the $a_1$ together with the values from the calculations before will of course reproduce the results from before and will give a good description of the data. To check whether a scenario with an explicit $a_1$ can also describe the data reasonably well, we have to demand that the coupling is not almost zero, and we expect the value of the coupling to be comparable to the values found in \cite{oseta1,meissnera1,pich}, i.e. comparable to Eq.(\ref{c1c2}). In order to get non zero couplings and still be reasonably close to the data, we have to keep the contribution from the WT term small. We start by choosing $\mu_1=\mu_2=M_\rho^2$ according to \cite{lutz2}, which we can expect to be a good choice by looking at Fig. \ref{rencomp1}. Scanning through the parameter space, it turned out that in most cases a two bump structure is observed. We show an example of this in Fig. \ref{a1} (set 3).
By finetuning the parameters, it is possible to merge these two bumps into one, which can also be seen in Fig. \ref{a1} (set 1 or set 2). The parameters leading to these curves are given in Tab. \ref{a1paras}. 
\begin{table}
\begin{center}
\begin{tabular}{|c||c|c|c|c|c|c|c|}
\hline
 & $M_{a_1}\,$[GeV] &  $f_A$ &  $c_1$ &  $c_2$ & $\mu_1\,[M_\rho^2$] & $\mu_2\,[M_\rho^2]$ & remark\\ \hline \hline
set 1 & 1.23  & $\frac{F_0}{\sqrt 2 M_\rho}\cdot 1.05$  & -$\frac{1}{4}\frac{1}{1.65}$ & $-\frac{1}{8}\frac{1}{1.6}$ & 2 & 1.05 &   \\ \hline
set 2 & 1.195  & $\frac{F_0}{\sqrt 2 M_\rho}\cdot 1.45$  & -$\frac{1}{4}\frac{1}{2.6}$ & $-\frac{1}{8}\frac{1}{1.6}$ & 1 & 2.5 &   \\ \hline
set 3 & 1.21 & $\frac{F_0}{\sqrt 2 M_\rho}\cdot 1.45$ & -$\frac{1}{4}\frac{1}{2.4}$ & $-\frac{1}{8}\frac{1}{1.6}$ & 1 & 2.5 &    \\ \hline
set 4 & 1.5 & $\frac{F_0}{\sqrt 2 M_\rho}$ & -$\frac{1}{4}$ & $-\frac{1}{8}$ & 2 & 5.5 & w/o WT \\ \hline
set 5 & 1.5 & $\frac{F_0}{\sqrt 2 M_\rho}$ & - & $-\frac{1.4}{8}$ & 2 & 5.8 & w/o WT, Eq.(\ref{oseta1vert})\\ \hline
set 6 & 1.2 & $\frac{F_0}{\sqrt 2 M_\rho}\cdot 1.05$ & -$\frac{1}{4}\frac{1}{1.7}$ & $-\frac{1}{8}\frac{1}{1.6}$ & 2 & 6 & $f_V,g_V$ Eq.(\ref{fvgvres})   \\ \hline
\end{tabular}
\end{center}
\caption{Different sets of parameters for the calculations with explicit $a_1$. The remark 'w/o WT' means that the WT term is not included, the additional remark 'Eq.(\ref{oseta1vert})' means that the $a_1$ decay vertex from Eq.(\ref{oseta1vert}) is used (which does not employ the parameter $c_1$) and '$f_V,g_V$ Eq.(\ref{fvgvres})' means that we choose the values from Eq.(\ref{fvgvres}) for $f_V$ and $g_V$.}\label{a1paras}
\end{table}
From set 1 and set 2, we see that there are different possible choices for the parameters, which can (more or less) describe the data. A deviation is only seen for $0.8\,\text{GeV}^2\lesssim s \lesssim 1.1\,\text{GeV}^2$.

\begin{figure}
\begin{center}
\includegraphics[width=9cm]{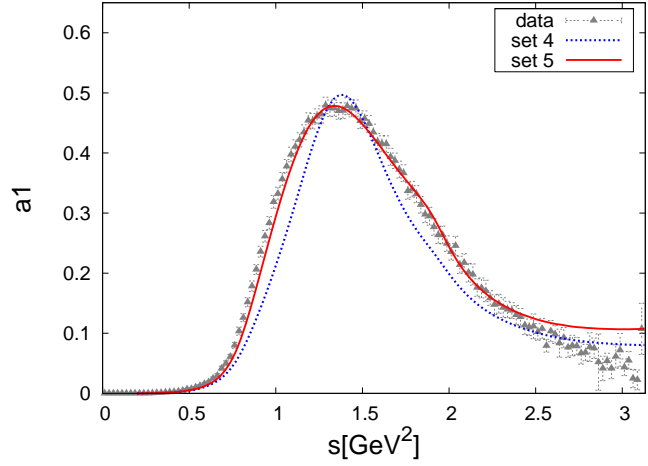}
\caption{Spectral function for the decay $\tau^-\rightarrow 2\pi^0 \pi^-\nu_\tau$ including the $a_1$ with different sets of paramaters in comparison to data from \cite{aleph1}. The WT was not included in these calculations. In addition, the curve labelled 'set 5' uses a different energy dependence to describe the $a_1$ decay (see text for details).}\label{a1wowt}
\end{center}
\end{figure}

Next we want to see, how far we get by switching off the WT term. We perform such an analysis without the WT term for completeness. We recall our strategy discussed in Section \ref{secunit} to approximate the $\phi V$ scattering kernel by possible resonances (here the $a_1$) plus contact terms of lowest order. In that strategy there is no justification to neglect even the lowest-order contact term, which is just the WT term with its strength fixed model independently by chiral symmetry breaking. In Fig. \ref{a1wowt} we see the spectral function calculated with explicit $a_1$ but without the WT term in the kernel (set 4). 
In that case the second bump disappears and by changing the parameters, one can determine the position and the height of the peak. Although one might expect that the width of the peak can be adjusted by the choice of $c_1$ and $c_2$, this is not the case, since there is a more complex interplay between the position and the width of the peak. There are lots of parameter choices which give a qualitatively similar curve. The curve labelled 'set 4' represents the best fit to the data by varying all parameters and we see that it agrees with the choice for $c_1$ and $c_2$ from \cite{meissnera1} and \cite{oseta1} (cf. Eq.(\ref{c1c2})). For the curve labelled 'set 5', we used the $a_1$ decay vertex from Eq.(\ref{oseta1vert}) and we see that in this case the shape of the peak is described very well. This shows the uncertainties in the shape of the width and how it is influenced by the energy dependence of the $a_1$ decay vertex. Therefore, we have to be careful to draw conclusions from the exact shape of the width.\\
From Fig. \ref{a1wowt} (set 4) and Fig. \ref{a1} one should not conclude that the WT term is a correction, which improves the shape of the width, since one does not simply switch on the WT term in order to get from Fig. \ref{a1wowt} to Fig. \ref{a1}. Instead the parameter sets are quite different (cf. Tab. \ref{a1paras}) and one has to finetune the parameters in order to arrive at a single reasonable peak, when one includes the WT term.\\
Looking at 'set 4' in Fig. \ref{a1wowt} one might worry that we do not have the most sophisticated model describing the explicit $a_1$ and that other models describe the data better (e.g. \cite{linsig,pich}). However, our description of the $a_1$ is completely sufficient to show the strong influence of the WT term.

\begin{figure}
\begin{center}
\includegraphics[width=9cm]{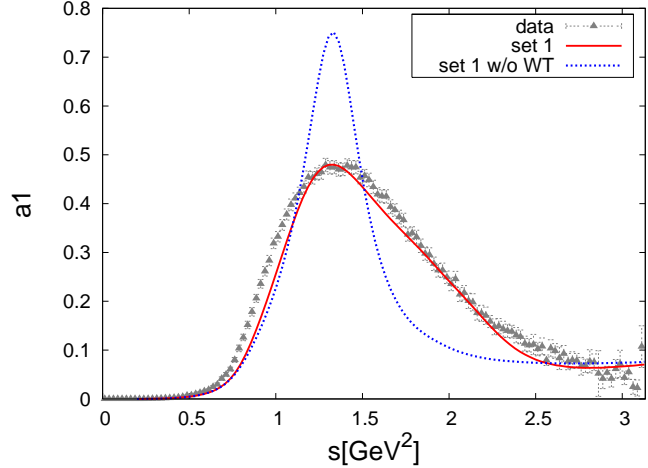}
\caption{Spectral function for the decay $\tau^-\rightarrow 2\pi^0 \pi^-\nu_\tau$ including the $a_1$ using paramater set 1 with and without including the WT term in comparison to data from \cite{aleph1}.}\label{a1wowt_2}
\end{center}
\end{figure}

Next, we want to see the role of the WT term played in the best result, which was shown in Fig. \ref{a1} (set 1). In Fig. \ref{a1wowt_2}
we plotted the result from parameter set 1 with and without including the WT term. Although with this choice of subtraction points the WT term is suppressed very strongly, it obviously still has a major influence on the result.

\begin{figure}
\begin{center}
\includegraphics[width=9cm]{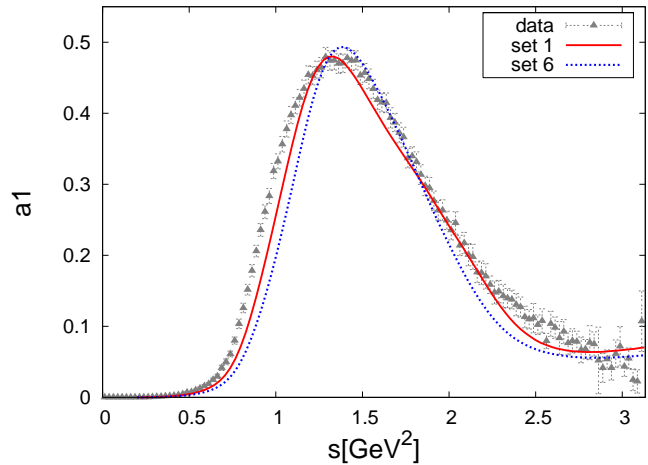}
\caption{Spectral function for the decay $\tau^-\rightarrow 2\pi^0 \pi^-\nu_\tau$ including the $a_1$ using different sets of paramaters in comparison to data from \cite{aleph1}.}\label{a1fvgv}
\end{center}
\end{figure}

In the discussion of the results without the explicit $a_1$, we showed results for the theoretically motivated values of $f_V$ and $g_V$ given in Eq.(\ref{fvgvres}). Using these theoretical values, it is possible to further suppress the WT term in comparison to the explicit $a_1$. In Fig. \ref{a1fvgv} 
we see the best result, which we found in this case (set 6). In Fig. \ref{gv} we found that the rise in the data for energies between about $0.7\,\text{GeV}^2 \lesssim s \lesssim 1.2\,\text{GeV}^2$ is described worse. Here, we find the same problem in the case of the $a_1$. However, we note that using the theoretically motivated parameters for $f_V$ and $g_V$ it is much easier to get rid of the second bump. In this case one can hardly call it finetuning to obtain a single reasonable one-peak structure. Nonetheless, we stress again that the $\tau$ decay data favour the experimental values for $f_V$ and $g_V$ given in Eq.(\ref{fvgvres1}) as compared to the theoretically motivated values of Eq.(\ref{fvgvres}). Using the latter the rise of the data in the energy region $0.7\,\text{GeV}^2 \lesssim s \lesssim 1.2\,\text{GeV}^2$ is underestimated (cf. Fig. \ref{gv} curves g1,g2 and Fig. \ref{a1fvgv} set 6).

Looking at the results including the $a_1$ it is not so easy to draw an immediate conclusion. For sure, one can say that the WT term has a major influence on the result. The second bump structure can be recovered in almost every calculation including the $a_1$ together with the WT term. An important point is that the inclusion of the WT term leads to very strong effects, although we already kept the contribution very small. Only by finetuning one can merge the two appearing bumps. However, merging two bumps by finetuning the parameters does not seem to be a natural way of reproducing the data. In other words: Why should an elementary state appear right at the position, where an attractive potential already created a peak? We note again that the strength of the WT term is model independently fixed by chiral symmetry breaking. Since the WT term alone already produces a peak at the right position, one could already expect that a description of the data including the $a_1$ has to be accompanied by a delicate choice of the parameters. Still, it would be too much to talk about a definite sign that there is no explicit $a_1$. However, the peculiarities with explicit $a_1$ together with the success of the description without the $a_1$ (molecule scenario) should be regarded as a good indication. In the next section, we will show that adding higher-order corrections to the WT term it is possible to systematically improve the situation in the molecule scenario and that the ordering of diagrams makes sense in this scenario without an explicit $a_1$.

\subsection{Higher-order terms}\label{sechores}

\begin{figure}
\begin{center}
\includegraphics[width=9cm]{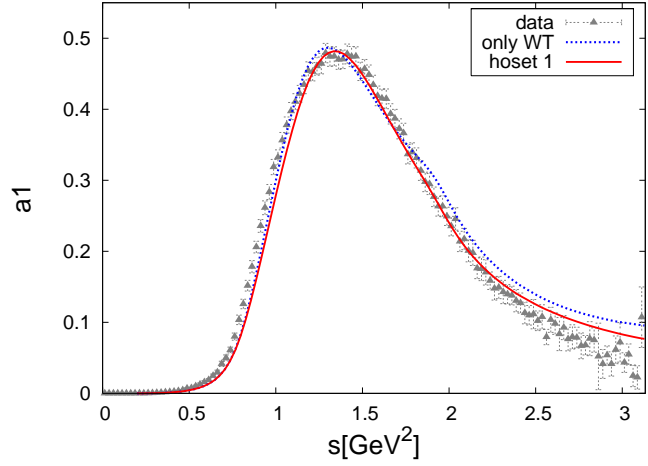}
\caption{Spectral function for the decay $\tau^-\rightarrow 2\pi^0 \pi^-\nu_\tau$ including higher-order terms in the kernel in comparison to data from \cite{aleph1}. For the choice of the parameters see Tab. \ref{tabho}.}\label{ho1}
\end{center}
\end{figure}

\begin{figure}
\begin{center}
\includegraphics[width=9cm]{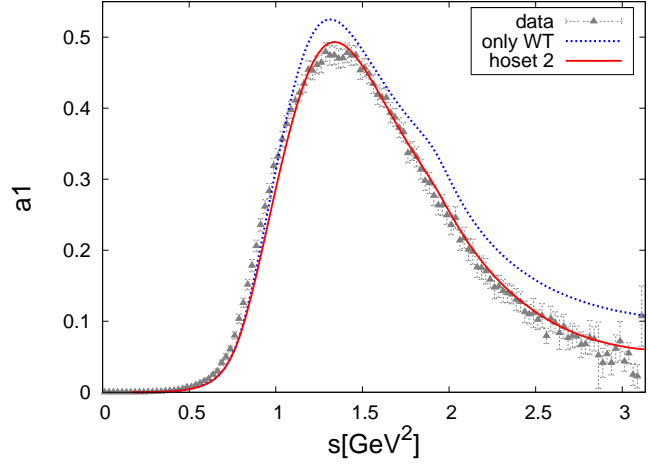}
\caption{Spectral function for the decay $\tau^-\rightarrow 2\pi^0 \pi^-\nu_\tau$ including higher-order terms in the kernel in comparison to data from \cite{aleph1}. For the choice of the parameters see Tab. \ref{tabho}. In the calculation including only the WT term we employ the same subtraction point $\mu_2=9\,M_\rho^2$ as in the calculation including the higher-order terms.}\label{ho2}
\end{center}
\end{figure}

\begin{figure}
\begin{center}
\includegraphics[width=9cm]{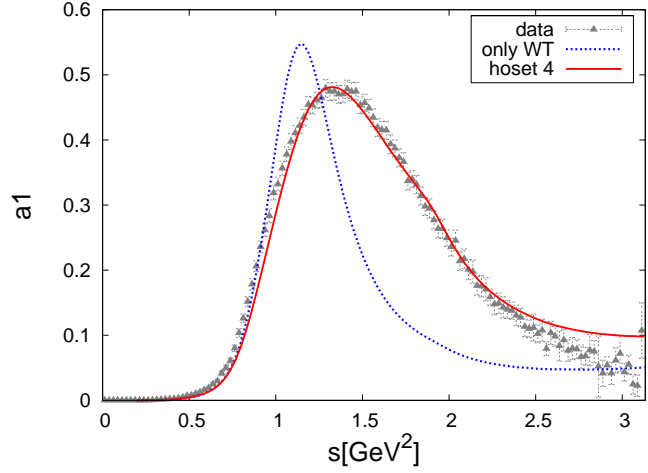}
\caption{Spectral function for the decay $\tau^-\rightarrow 2\pi^0 \pi^-\nu_\tau$ with and without higher-order terms using $\mu_1=\mu_2=8.5M_\rho^2$ in comparison to data. The parameters for the higher-order corrections are given in Tab. \ref{tabho}.}\label{ho3}
\end{center}
\end{figure}

In Section \ref{sechig} we determined the corrections to the kernel at $\mathcal O(q^2)$, which led to six new unknown parameters. In the following we leave out the explicit $a_1$ again and show the influence of these corrections on the results. In Fig. \ref{ho1} we show the spectral function with and without including the higher-order correction. There are several parameter sets (Tab. \ref{tabho}), which can describe the data in a qualitatively similar way. We see that the higher-order terms can be chosen such that they systematically improve the agreement with the data. Note that the size of the higher-order terms is not constrained by chiral symmetry (except that they should be of natural size - a demand of every effective field theory).
\begin{table}
\begin{center}
\begin{tabular}{|c||c|c|c|c|c|c|c|c|}
\hline
& $\;\;\lambda_1\;\;$ &  $\lambda_2\, [\text{GeV}^{-1}]$ &  $\;\;\lambda_3\;\;$ &  $\;\;\lambda_4\;\;$ &  $\lambda_5\, [\text{GeV}^{-1}]$ &  $\;\;\lambda_6\;\;$ & $\mu_1 \,[\text{GeV}^2]$ & $\mu_2 \,[\text{GeV}^2]$ \\ \hline \hline
hoset 1 & 0 & 0 & 1.5 & -1.4 & 0 & 0 & $M_\rho^2$ & $8.5M_\rho^2$\\ \hline
hoset 2 & 0.6 & 0.3 & 2.5 & 0 & 0 & 0 & $M_\rho^2$ & $9M_\rho^2$\\ \hline
hoset 3 & 0 & -0.3 & 0 & -1.4 & 0 & 0 & $M_\rho^2$ & $8.5M_\rho^2$\\ \hline
hoset 4 & 0.85 & 0 & 0 & -0.45 & 0 & 0 & $8.5M_\rho^2$ & $8.5M_\rho^2$\\ \hline
\end{tabular}
\end{center}
\caption{Different sets of parameters, which yield a good description of the spectral function.}\label{tabho}
\end{table}\\
Next we investigate the connection between the higher-order terms and the subtraction points. Changing $\mu_2$ can hardly be compensated by the higher-order terms, which is expected, since $\mu_2$ acts as a higher-oder correction to the $W$ decay vertex and not to the scattering amplitude. However, a slight raise in $\mu_2$ can be compensated, as can be seen in Fig. \ref{ho2}. There we use $\mu_2=9M_\rho^2$ for both calculations. We recall that without higher-order terms the best value was $\mu_2=8.5M_\rho^2$ (cf. Fig. \ref{sub2p}).
It can also be seen that the higher-order terms do not touch the energy region up to about $s\approx 1\,$GeV$^2$.\\
Next we want to discuss the connection to $\mu_1$. Here we can expect that the higher-order terms are at least to some extent able to compensate for changes. In Fig. \ref{ho3}
we can see that we can account for moving the subtraction point $\mu_1$ by changing the parameters of the higher-order terms. The parameters are again given in Tab. \ref{tabho}. We see that the compensation is even better than expected, since the corrections up to order $\mathcal O(q^2)$ do not carry all the structures, which might be influenced by moving the subtraction point.\\
It is also interesting to look directly at the changes induced in the scattering amplitude.
\begin{figure}
\begin{center}
\begin{tabular}{cc}
\includegraphics[width=7cm]{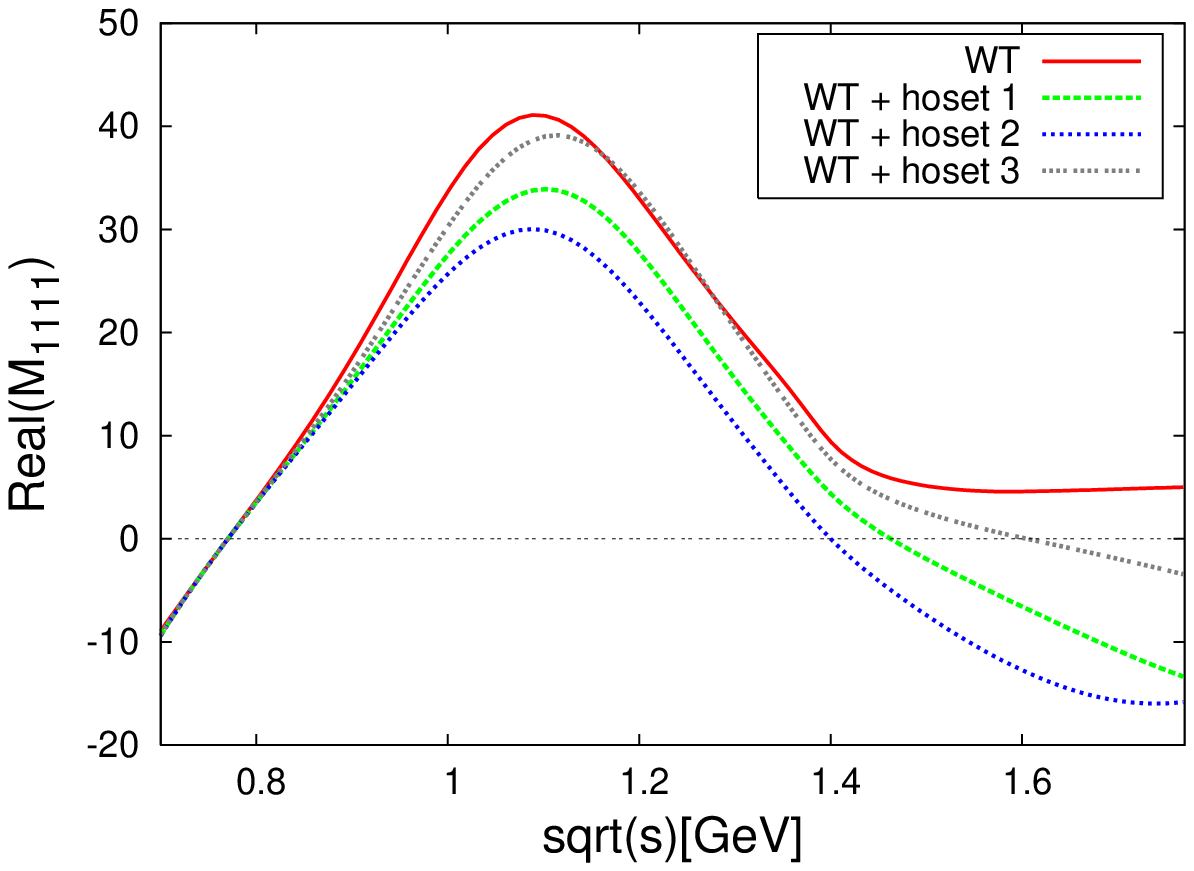} & \includegraphics[width=7cm]{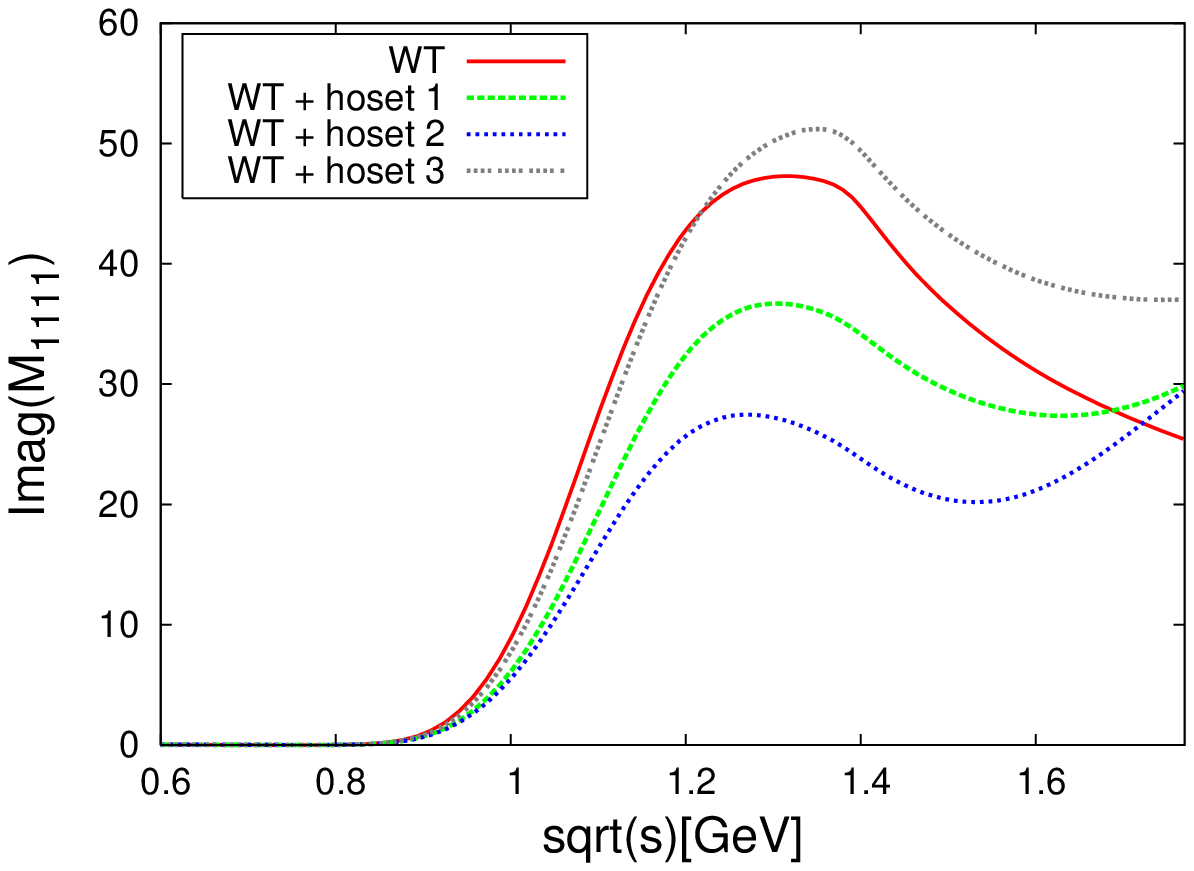} \\
(a) & (b) \\
\includegraphics[width=7cm]{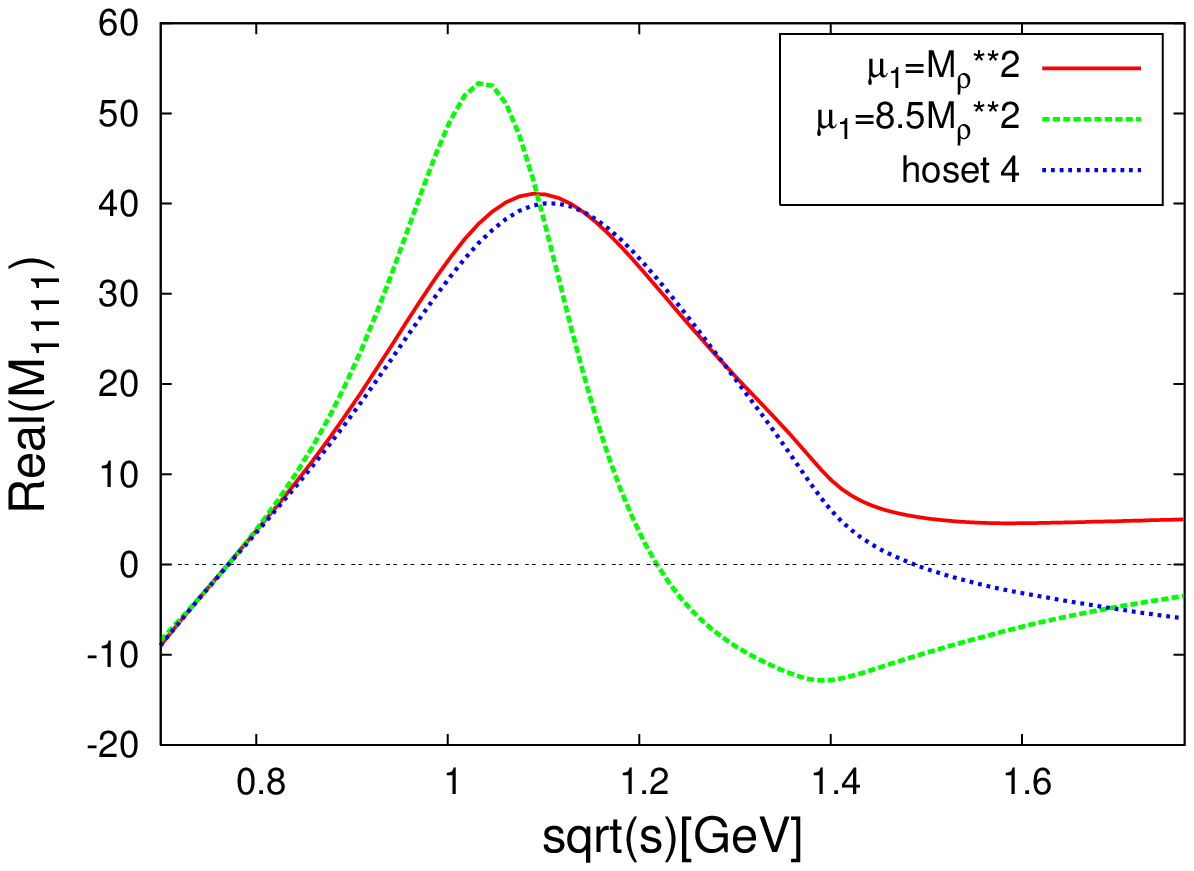} & \includegraphics[width=7cm]{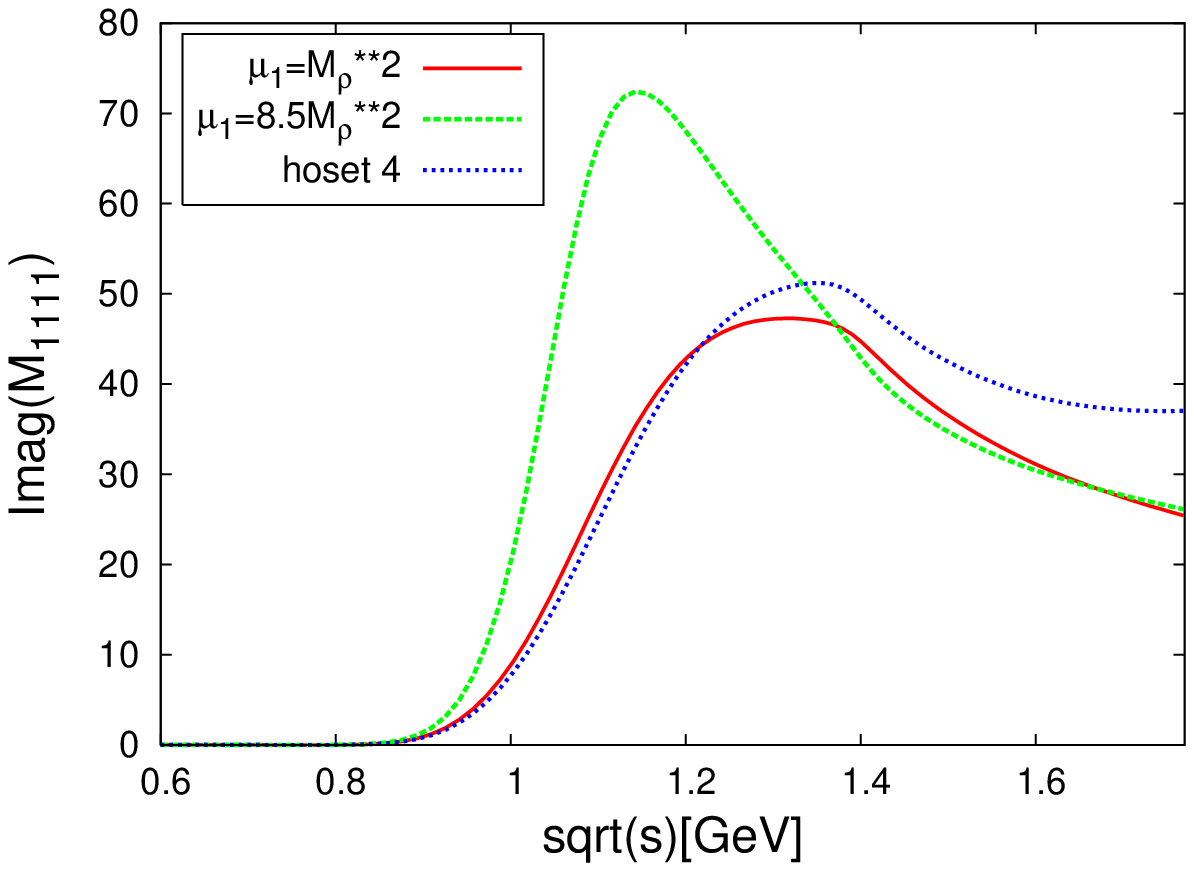} \\
(c) & (d) \\
\end{tabular}
\caption{Real (left) and imaginary (right) part of the scattering amplitude for $\pi\rho$ scattering with and without higher-order corrections. The upper two plots show the scattering amplitudes for hoset 1-3 in comparison to a calculation without higher-order terms and $\mu_1=M_\rho^2$. The lower two plots show the scattering amplitude for hoset 4 in comparison to a calculation without higher-order corrections and $\mu_1=M_\rho^2$ and $\mu_1=8.5M_\rho^2$, respectively.}\label{homatel}
\end{center}
\end{figure}
In Fig. \ref{homatel} we see the real and imaginary part of the scattering amplitude for the different parameter sets given in Tab. \ref{tabho} in comparison to the calculation without higher-order corrections. In order not to overload the figures, we show four different plots. Figs. \ref{homatel}(a) and (b) show the scattering amplitudes for the first three parameter sets in comparison to a calculation without higher-order corrections and $\mu_1=M_\rho^2$. We see that the scattering amplitude is not modified much. Only 'hoset 2' shows a different structure, which becomes most obvious in the imaginary part. This change in the scattering amplitude seems to be correlated with the parameter $\lambda_3$, which we will further investigate, when we look at the Dalitz plot projections in Section \ref{secdal}. Figs. \ref{homatel}(c) and (d) show the scattering amplitudes for 'hoset 4', which was chosen to compensate for the change in the subtraction point $\mu_1$. The figure shows the scattering amplitude for $\mu_1=M_\rho^2$ and $\mu_1=8.5 M_\rho^2$ without higher-order terms in comparison to $\mu_1=8.5M_\rho^2$ with higher-order corrections. We see that the corrections bring the scattering amplitudes for $\mu_1=8.5M_\rho^2$ back into the shape they had before, when we used $\mu_1=M_\rho^2$ without higher-order corrections. In other words, changes in the renormalisation point can be replaced by changes in the higher-order terms. The renormalisation scale dependence is reduced as it should, when including higher-order terms.

One might argue that including the higher-order corrections was unnecessary and describing the data with 6 parameters is no success. However, the point in including the higher-order corrections is not that we can describe the data with seven parameters, but that they systematically improve the result. In case of the inclusion of an explicit $a_1$, we saw that adding the WT term to the $a_1$ interaction worsened the results. Here, however, adding the higher-order terms to the kernel behaves as a correction. Note that the calculations are not ordered according to usual perturbation theory. Instead the kernel of the Bethe-Salpeter equation is calculated in perturbation theory. The convergence of that kind of perturbative expansion is not guaranteed. Therefore it is encouraging to see that the next to leading order terms behave as a correction and are even able to improve the agreement with the data.\\
We note that there are many possible choices for the parameters, which describe the data. In Section \ref{secdal} we will see that the different sets can be further discriminated by looking at the Dalitz plot projections.

\subsection{Dalitz plot projections}\label{secdal}

\begin{figure}
\begin{center}
\includegraphics[width=14cm]{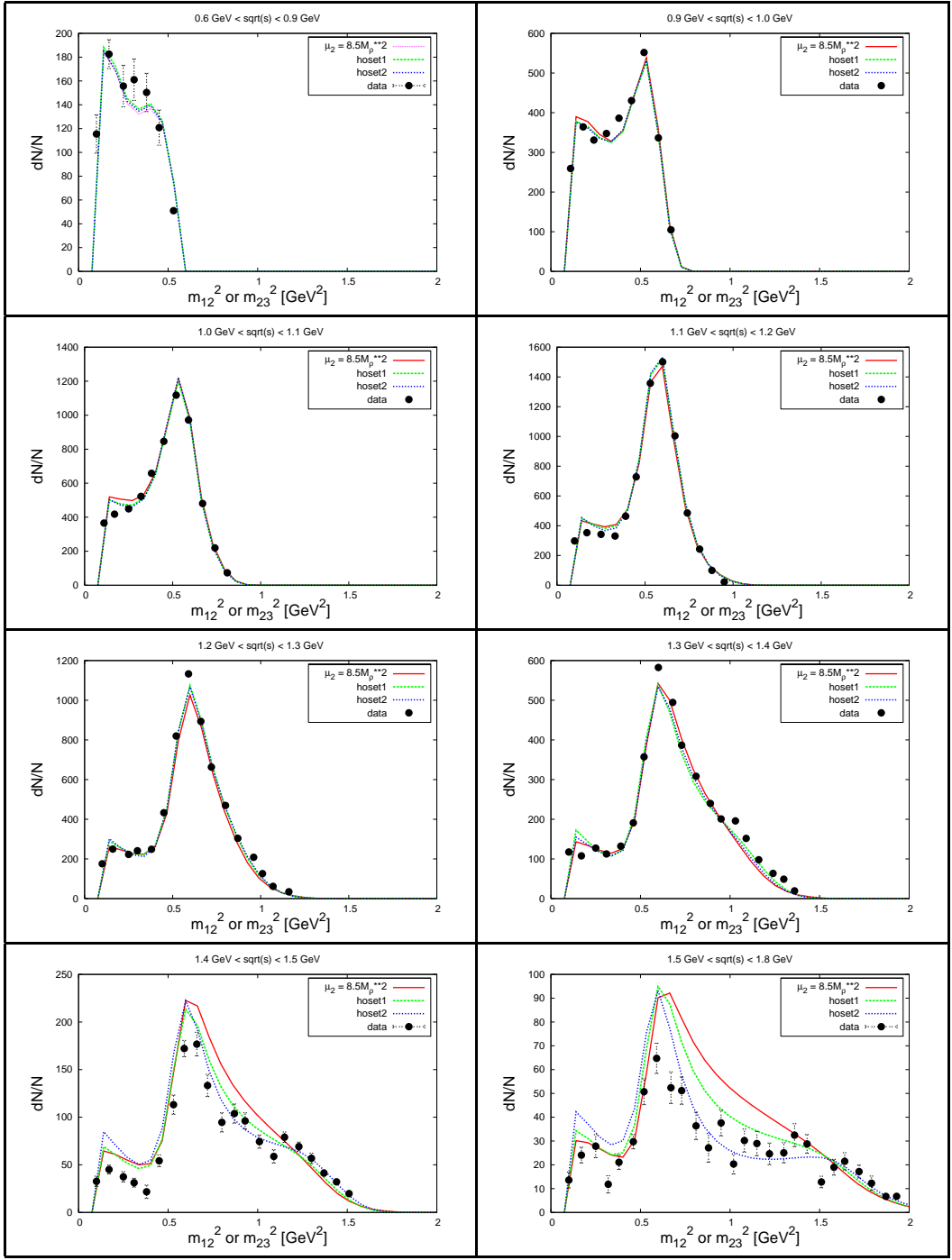}
\caption{Dalitz plot projections in $m_{12}^2$ or $m_{23}^2$ with and without higher-order corrections in comparison to data from \cite{dalitz}. The different parameter sets can be found in Tab. \ref{tabho}. The curve labelled $\mu_2=8.5M_\rho^2$ corresponds to a calculation using the WT term only, $\mu_1=M_\rho^2$ and $\mu_2=8.5M_\rho^2$.}\label{dalitz1}
\end{center}
\end{figure}

\begin{figure}
\begin{center}
\includegraphics[width=14cm]{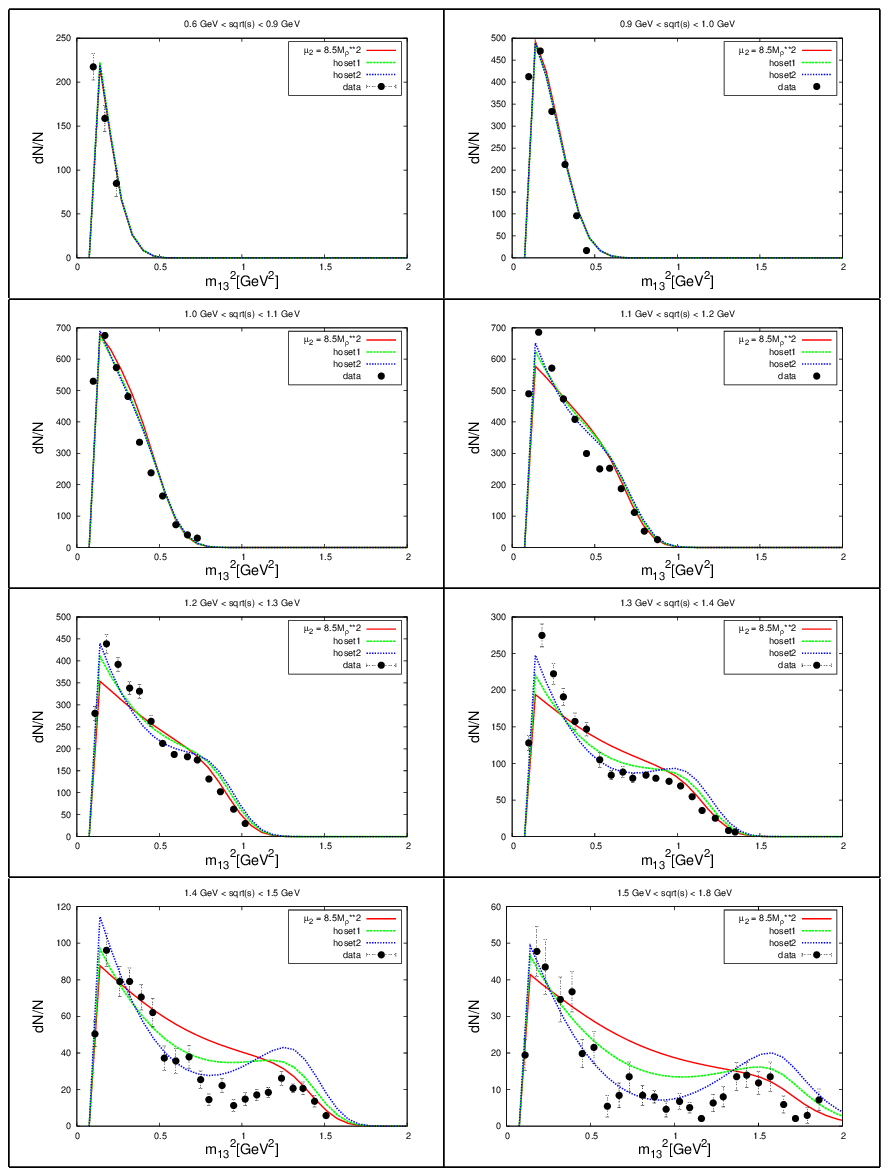}
\caption{Dalitz plot projections in $m_{13}^2$ with and without higher-order corrections in comparison to data from \cite{dalitz}. The different parameter sets can be found in Tab. \ref{tabho}. The curve labelled $\mu_2=8.5M_\rho^2$ corresponds to a calculation using the WT term only, $\mu_1=M_\rho^2$ and $\mu_2=8.5M_\rho^2$.}\label{dalitz2}
\end{center}
\end{figure}

\begin{figure}
\begin{center}
\begin{tabular}{cc}
\includegraphics[width=7cm]{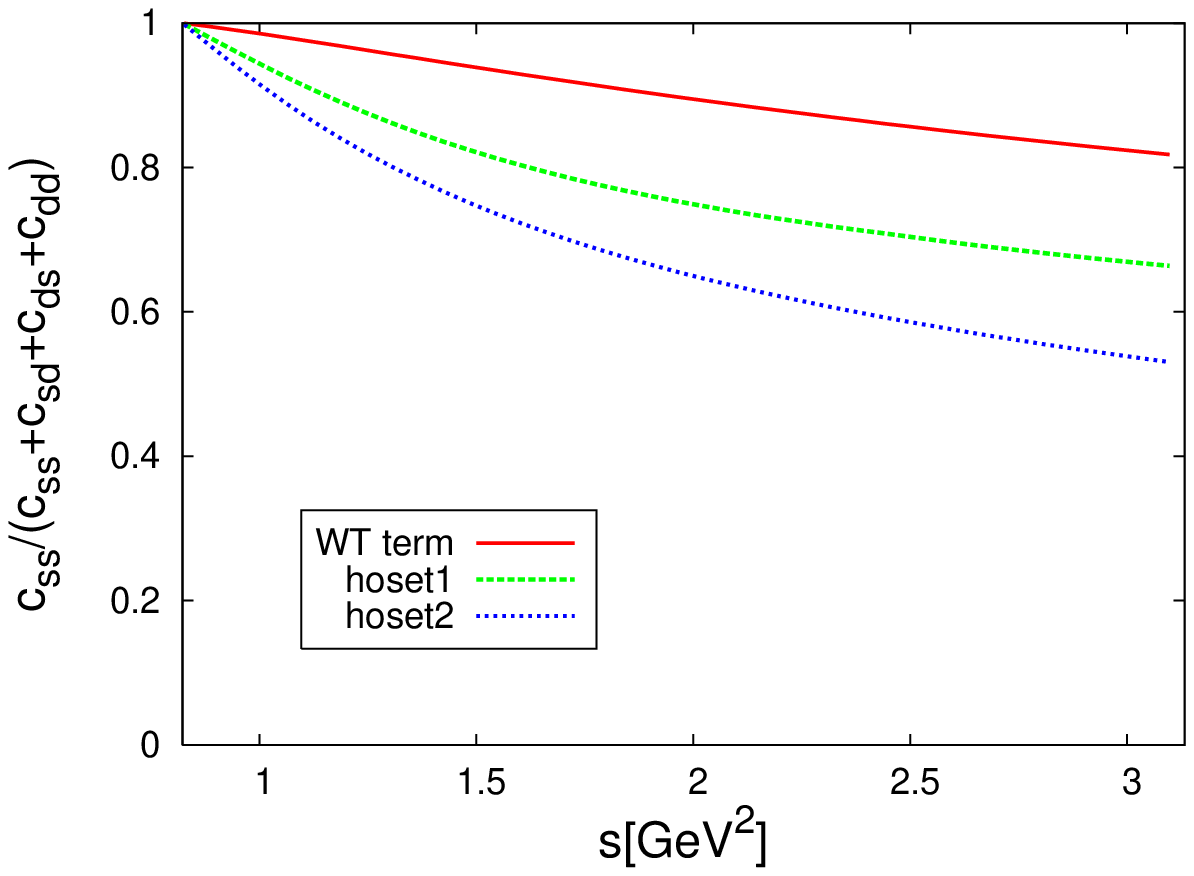} & \includegraphics[width=7cm]{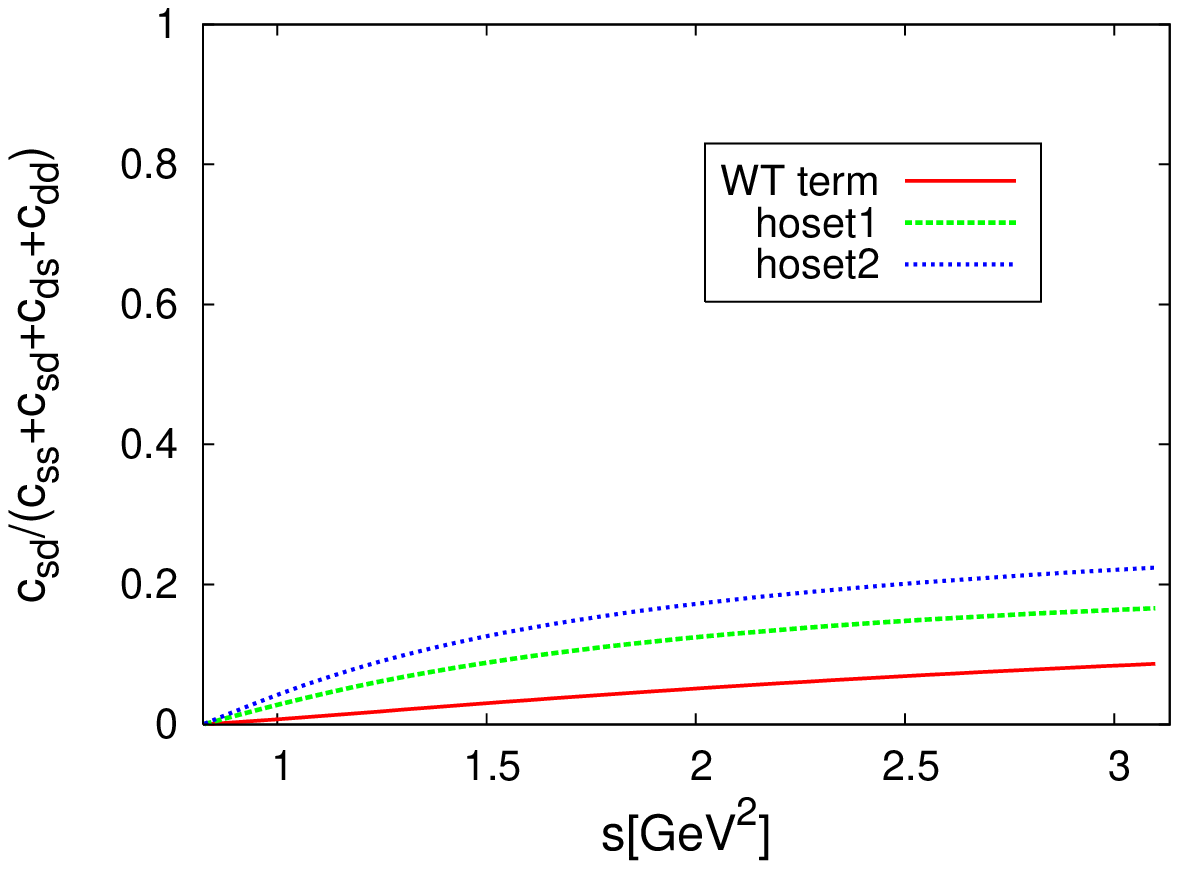}
\end{tabular}
\caption{Population of angular momentum states expressed in $c_{ss}/(c_{ss}+2c_{sd}+c_{dd})$ (left) and $c_{sd}/(c_{ss}+2c_{sd}+c_{dd})$ (right). The curve labelled 'WT term' shows the population for the WT term only, whereas the other kernels include higher-order corrections with the parameters given in Tab. \ref{tabho}.}\label{sdwave}
\end{center}
\end{figure}

In Fig. \ref{dalitz1} we show the Dalitz plot projections in $m_{12}^2$ or $m_{23}^2$ for the calculation using the WT term only and for the calculation including the higher-order terms in comparison to data from \cite{dalitz}. We determined the normalisation of the theoretical curve such that the area under all curves, corresponding to different slices of $\sqrt s$, agrees with the area under the data points. We also subtracted the contribution to the data, which was identified as background in \cite{dalitz}. Using this normalisation, we do not really lose any information, since we already saw before that the spectral function was well reproduced for all invariant masses, which implies a proper total decay width and therefore a proper normalisation. Fig. \ref{dalitz1} clearly shows that the final state is dominated by the $\rho$ meson, and the data are described quite well by all parameter sets. 
The last two plots seem to show an improvement by including the higher-order corrections. The improvement in the last two plots is most pronounced for 'hoset 2'. But since we overshoot at lower $m_{12}^2$ and the error bars are pretty large for these invariant masses, the advantage is not very stringent.
We recall that $q_1$ and $q_3$ are the momenta of the likewise pions and that the amplitude is symmetric under the exchange $q_1\rightarrow q_3$. Thus, $m_{12}^2$ and $m_{23}^2$ are the invariant mass of the intermediate $\rho$, which we clearly see in the Dalitz plot projections.\\
In Fig. \ref{dalitz2} we plot the number of decays versus $m_{13}^2$, which is the invariant mass of the likewise pions. These pions do not build up the $\rho$ and therefore the structure in these plots is completely different from Fig. \ref{dalitz1}. When we look at Fig. \ref{dalitz2}, we also find that the calculations including the higher-order corrections describe the data better, which again is more pronounced for 'hoset 2'. The steep rise at small $m_{13}^2$ is much better reproduced by 'hoset 2' and also the additional structure at higher invariant masses is reproduced better with 'hoset 2', although we overshoot that structure.
We note that using 'hoset 3' and 'hoset 4', we do not get a noteworthy difference in the Dalitz plot projections in comparison to a calculation without higher-order corrections. Thus, the improvement in the Dalitz plots seems to be correlated with the parameter $\lambda_3$, which is non-vanishing for 'hoset 1' and 'hoset 2' (cf. Tab. \ref{tabho}).\\
It is interesting to see the amount of $d$-wave contributions from the different parameter sets. In Fig. \ref{sdwave}
we plotted the absolute value of the ratio of the respective coefficient $c_{ss}$ and $c_{ds}$ to the sum of all coefficients (see Appendix \ref{orbsec}). We see, that the parameter sets, which describe the Dalitz plot projections better, clearly have a higher $d$-wave contribution. Thus, our calculation indicates a population of $d$-waves in the $\tau$ decay in the amount, which is shown in Fig. \ref{sdwave}. We note that the leading order contribution to $c_{sd}$ is given by the terms proportional to $\lambda_3$ and $\lambda_6$ in Eq.(\ref{hokern}). A statement about pure $d$-wave transitions given by $c_{dd}$ would be more complicated, since terms of higher chiral order than $q^2$ would contribute at leading order.

\section{Summary and Outlook}\label{secsum}

We calculated the process $\tau^-\rightarrow 2\pi^0\pi^-\nu_\tau$ for different scenarios. In the first scenario (molecule scenario) we analysed the decay based on the recently developed techniques to generate axial vector resonances dynamically \cite{lutz2,osetaxial}. The picture we promote is that the process is dominated by $\pi\rho$ final state interactions, which are described by iterating the WT term. The weak decay is part of the standard model and the WT term is predicted parameter free from chiral symmetry. The remaining coupling constants ($f_V,g_V$), which describe the interaction of the vector mesons are determined by the properties of the $\rho$ \cite{vecrep}. The only unknown parameters in the calculation enter through the renormalisation of the loop integrals. We introduced two subtraction constants to render the loop integrals finite. One subtraction constant ($\mu_1$) renormalises the loops in the scattering amplitude describing the final state interactions. This parameter was already introduced in \cite{lutz2} and fixed by crossing symmetry arguments. The other subtraction constant ($\mu_2$) enters in the renormalisation of the first loop, which contains the $W$ decay vertex. We investigated the influence of these parameters on the results. First, we varied $\mu_1$ and $\mu_2$ simultaneously with $\mu_1=\mu_2$ and compared in particular the different values used in \cite{lutz2} and \cite{osetaxial}. We found that all choices produce a peak at the same position with a different height. The position of this peak was roughly in the region of the resonant structure seen in the data, but the width always turned out to be too small. Afterwards we investigated the influence of $\mu_2$ by keeping $\mu_1$ fixed. Using the crossing symmetry argument from \cite{lutz2} in order to determine $\mu_1$, leaves us with one free parameter. Fitting this parameter $\mu_2$, we reproduced the spectral function for the decay $\tau^-\rightarrow 2\pi^0\pi^-\nu_\tau$ quite well.\\
In a second scenario, we explicitly introduced the $a_1$ in the calculation. This introduces new parameters, namely the mass of the $a_1$, its coupling $f_A$ to the $W$ boson and the couplings $c_1$ and $c_2$ to the vector-meson Goldstone boson states. The most obvious feature in that calculation is, that due to the strong influence of the WT term a second bump appears. Finetuning the parameters, one can merge the two bumps into one and the data can be described more or less satisfyingly. However, the results of these calculations are unsatisfying. An important point is that the inclusion of the WT term leads to very strong effects, although we already kept the contribution very small. Merging two bumps by finetuning the parameters does not seem to be a natural way of reproducing the data. Since the WT alone already produces a peak at the right position, one could expect already that a description of the data including the $a_1$ has to be accompanied by a delicate choice of the parameters. In addition, one can obviously not talk about a small correction, which is induced by the WT term.\\
A further improvement of the molecule scenario was found by introducing higher order corrections to the kernel. These corrections introduce six new parameters and many combinations of these parameters could be found, which fit the spectral function very well. The correction induced by these terms were well behaved. That was not clear from the beginning, since we calculated the kernel of the Bethe-Salpeter equation perturbatively, which does not automatically guarantee that we picked up all important contributions for the scattering amplitude itself. Therefore, this is an encouraging fact, which puts further foundation to the calculation and shows its systematic nature.\\
Comparing our calculation to the Dalitz projections, we found that including higher order terms, which carry $d$-wave components, describe the data better. We compared the size of the coefficient $c_{sd}$, which describes the transitions from an $s$-wave to $d$-wave state, for different parameter sets. The size of this coefficient was clearly correlated with the qualitative description of the Dalitz plot data.\\
To summarise, one finds that without the explicit $a_1$ one has a well behaved model, which can be systematically improved and which describes the data very well. Most parameters (in the simplest scenario, all but one) are fixed by chiral symmetry breaking and the well known properties of the $\rho$. Including an explicit $a_1$ leads to peculiar properties, if one tries to generate the width consistently from the Bethe-Salpeter equation and includes the WT term. When we tried to describe the data with an explicit $a_1$ the strength of the WT interaction caused the most severe problems. On the other hand, this strength is fixed by chiral symmetry breaking. In addition, we recall that taking into account an explicit $a_1$ and the WT interaction is not double counting. Essentially we claim that the WT interaction should not be disregarded as has, however, been done in many previous approaches. On the other hand without an explicit $a_1$ the WT interaction has the right strength to generate a resonant structure dynamically. These indications point towards a dynamical nature of the $a_1$ as a (coupled-channel) meson molecule.\\
As an outlook we note that a further step in the calculation would be to include medium effects in order to see what happens to the $a_1$ in case we approach the chiral symmetry restoration \cite{rapp}. In principle, when the restoration happens, the axial-vector spectral function, defined in Section \ref{secwidth}, must be degenerate with the corresponding vector spectral function. In the latter the $\rho$ meson prominently appears, at least in the vacuum \cite{aleph1}. It is, however, not so clear what chiral restoration implies for the specific part of the spectral function with a three-pion final state. In any case one would expect a drastic reshaping of both the vector and the axial-vector spectral function.\\
It would also be interesting to figure out how well the molecule scenario agrees with QCD lattice calculations \cite{lattice} of the axial-vector current-current correlator (in the specific region accessible by lattice QCD). Here one has to perform the calculations with a higher pion mass in order to connect to lattice QCD calculations. This also brings into play pion mass corrections to the involved coupling constants as for example $F_0,f_V,g_V$ \cite{stefan2}. Still one can expect that the presented framework offers enough predictive power to obtain a valuable comparison to lattice QCD.

\begin{acknowledgments}
We would like to thank M. F. M. Lutz and E. Kolomeitsev for useful discussions. We would also like to acknowledge stimulating discussions with E. Oset and D. Rischke. In addition, we thank Julian Hofmann for discussions on the structure of the higher order terms. We also thank Hasko Stenzel for clarifying our questions concerning the data and we are grateful to U. Mosel for stimulating discussions and continuous support. This work was supported by DFG and GSI.
\end{acknowledgments}

\appendix

\section{Projectors}\label{secproj}

We briefly summarise the most important formulas, which define the projectors and which are used to determine the expansion coefficients. A scalar amplitude for the scattering of vector mesons with a scalar particle can be expanded as follows \cite{wick}
\begin{equation}\label{heli}
\epsilon_\mu^\dagger(\overline p, \overline\lambda)T^{\mu\nu}\epsilon_\nu(p,\lambda) = \sum_{JM} \frac{2J+1}{4\pi} D_{M\overline\lambda}^{J\ast}(\overline\phi,\overline\theta,-\overline\phi)D_{M\lambda}^J(\phi,\theta,-\phi) \langle JM\overline\lambda | T | JM\lambda\rangle\,,
\end{equation}
where $p,\lambda\,(\overline p,\overline\lambda)$ are the momenta and helicities of the incoming (outgoing) vector mesons, $D$ are the Wigner rotation functions and $| JM\lambda\rangle$ denotes a state with total angular momentum $J$, its projection $M$ and with the helicity of the vector particle being $\lambda$. Choosing the incoming particles to fly along the $z$-axis and the scattered particles to move in the $xz$ plane, the formula  reduces to
\begin{equation}\label{heli2}
\epsilon_\mu^\dagger(\overline p, \overline\lambda)T^{\mu\nu}\epsilon_\nu(p,\lambda)=\sum_J (2J+1)\langle\overline \lambda|T^J|\lambda \rangle d_{\lambda\overline\lambda}^J(\theta)\,,
\end{equation}
where we omitted $J,M$ in the denotation of the states and $d$ are the Simplified Wigner functions or $d$-functions. By Lorentz invariance the scattering amplitude can be written in terms of five scalar functions $F_i$
\begin{equation}\label{lor}
T_{\mu\nu}= \sum_i F_i L_{\mu\nu}^i\,,
\end{equation} 
with
\begin{equation}\label{thels}
\begin{split}
&L_1^{\mu\nu}=g^{\mu\nu}-\frac{w^\mu w^\nu}{s},\quad L_2^{\mu\nu}= w^\mu w^\nu, \quad L_3^{\mu\nu}=w^\mu \overline q^\nu - w^\mu w^\nu\frac{\overline q\cdot w}{s}\\
&L_4^{\mu\nu}=q^\mu w^\nu - w^\mu w^\nu\frac{q\cdot w}{s}, \quad L_5^{\mu\nu}=\left(q^\mu - w^\mu\frac{q\cdot w}{s}\right)\left(\overline q^\nu - w^\nu\frac{\overline q\cdot w}{s}\right)\,,
\end{split}
\end{equation} 
where $q\,(\overline q)$ is the incoming (outgoing) momentum of the Goldstone boson. We note that there are only five independent terms, since terms containing $p^\nu$ or $\overline p^\mu$ vanish due to $\epsilon^\mu(p)p_\mu=0$. Thus, using the orthogonality relation of the $d$-functions, one can express the expansion coefficients of Eq.(\ref{heli2}) in terms of the $F_i$
\begin{equation}\label{exp}
\langle \overline \lambda |T^J|\lambda\rangle =\frac{1}{2}\int_0^\pi\epsilon_\mu^\dagger(\overline p,\overline  \lambda)\left(\sum_i F_i L_i^{\mu\nu}\right)\epsilon_\nu(p,\lambda)d_{\lambda\overline\lambda}^J \sin\theta d\theta \,.
\end{equation} 
We further introduce parity eigenstates, which are given by
\begin{equation}
\langle 1_\pm | = \frac{1}{\sqrt 2} (\langle -1| \pm \langle 1|)\,.
\end{equation}
The defining equation for a projector with total angular momentum $J$, its projection $M$, parity $P$ and helicities $\lambda_2,\lambda_3$ is
\begin{equation}\label{projprop}
\epsilon^{\mu\dagger}(\overline p,\lambda_1) Y_{\lambda_2 \lambda_3\mu\nu}^{JM^P}(\overline q,l,s) \epsilon^\nu(l,\lambda_4) = \delta_{|\lambda_1|\lambda_2}\delta_{\lambda_3|\lambda_4|}(2J+1) D^{\ast J}_{M\lambda_1}(\overline\Omega) D^{J}_{M\lambda_4}(\Omega)\left(\frac{1}{\sqrt 2}\right)^{\lambda_2 + \lambda_3} P^{(\lambda_1 - \lambda_4)/2}\,.
\end{equation}
For $J^P=1^+$ and the kinematics described above the explicit form of the projectors is
\begin{equation}\label{eqproj}
\begin{split}
Y_{11\mu\nu}^{1^+}&=\frac{3}{2}\left(-L^1_{\mu\nu}+L^2_{\mu\nu}\frac{\omega\overline\omega x}{p \overline p s}+L^3_{\mu\nu}\frac{-\overline \omega}{\overline p^2 \sqrt s}+ L^4_{\mu\nu}\frac{-\omega}{p^2\sqrt s}\right)\\
Y_{10\mu\nu}^{1^+}&=M\frac{3}{\sqrt 2}\left(-\frac{\overline \omega x}{p\overline p s} L_{\mu\nu}^2 + L_{\mu\nu}^4 \frac{1}{p^2\sqrt s}\right)\\
Y_{01\mu\nu}^{1^+}&=-\overline M\frac{3}{\sqrt 2}\left(\frac{\omega x}{p \overline p s} L_{\mu\nu}^2 - L_{\mu\nu}^3 \frac{1}{\overline p^2 \sqrt s}\right)\\
Y_{00\mu\nu}^{1^+}&= \frac{3 M\overline M x L^2_{\mu\nu}}{p\overline p s}\,,
\end{split}
\end{equation}
where $M,\omega\,(\overline M,\overline\omega)$ are the mass and the energy of the incoming (outgoing) vector meson, $p_{cm}$ the centre-of-mass momentum and $s=(p+q)^2$ the total invariant energy of the process. For practical calculations in the centre-of-mass system, however, it is enough to know Eq.(\ref{projprop}) and the expansion coefficients from  Eq.(\ref{exp}).

\section{Connection between helicity states and orbital angular momentum}\label{orbsec}

In order to determine the $s$- and $d$-wave component of the vector-meson Goldstone boson two-particle state, we need to know the relation between the helicity states and the orbital angular momentum $l$. In particular, we first want to determine the following overlap
\begin{equation}\label{conneq1}
\langle J,M;l,s=1|J,M,\lambda\rangle = \,?
\end{equation}
In order to do so, we express both states in Eq.(\ref{conneq1}) in terms of orbital angular momentum and spin states, which is pretty simple for the left hand side. The states of total angular momentum $J$ can be written as a combination of states with definite orbital angular momentum $l$ and spin $s$
\begin{equation}\label{coneq1}
|J,M;l,s\rangle = \sum_{m_s} C(mm_s(ls)JM) |l,m\rangle |s,m_s\rangle\,,
\end{equation}
where $s=1$ is the spin of the vector meson, $m_s$ the $z$-projection of the spin, $m$ the z-projection of the orbital angular momentum, $M=m+m_s$ and $C$ is a Clebsch-Gordan coefficient. We choose the following notation for the Clebsch-Gordan coefficients
\begin{equation}
\langle j_1j_2 ,m_1 m_2|j_1j_2,jm\rangle = C(m_1m_2(j_1j_2)jm)\delta_{m,m_1+m_2}\,.
\end{equation}
Next we turn to the state $|J,M,\lambda\rangle$ in Eq.(\ref{conneq1}). We want to express the helicity states of the moving system in terms of the spin and orbital angular momentum states. Since the spin and the orbital angular momentum are not conserved quantum numbers in a relativistic framework, the helicity states will be a mixture of different states. We need the following relations
\begin{equation}
|l,m\rangle = \sqrt{\frac{2l+1}{4\pi}}\int d\Omega |\theta,\phi\rangle D^{l\ast}_{m,0}(\phi,\theta,0)\,,
\end{equation}
and the inverse of that equation, which is
\begin{equation}
|\theta,\phi\rangle = \sum_{l,m} |l,m\rangle \sqrt{\frac{2l+1}{4\pi}} D^l_{m0}(\phi,\theta,0)\,.
\end{equation}
At the same time we can write for a helicity state moving along the $z$ axis $|\hat z,\lambda\rangle$
\begin{equation}\label{peq}
|\hat z,\lambda\rangle = |\hat z\rangle |s,\lambda\rangle\,,
\end{equation}
where $|s,\lambda\rangle$ is the usual spin state with $m_s=\lambda$. Although the spin is not a conserved quantum number, it coincides with the helicity state in the rest frame of the particle. Since one can not produce any orbital angular momentum along the direction of motion, after a boost the $z$ projection of the total angular momentum is still given by the spin projection of the particle, which is the same as the helicity. Thus, we can use the above decomposition. Next we apply the rotation operator $U(\phi,\theta,0)$ to the state. After the rotation the spin and helicity states will not be the same anymore, but the connection is given by the Wigner rotation functions. We have to rotate each factor on the right hand side of Eq.(\ref{peq}) separately, which gives
\begin{equation}
|\theta,\phi,\lambda\rangle = U(\phi,\theta,0)|\hat z,\lambda\rangle = \sum_{m_S}|\theta,\phi\rangle D^1_{m_S\lambda}(\phi,\theta,0)|1,m_S\rangle\,.
\end{equation}
Applying the projection operator (see \cite{wuki} or \cite{wick}) on definite total angular momentum states, we get
\begin{equation}
\begin{split}
&|J,M,\lambda\rangle = \sqrt{\frac{2J+1}{4\pi}}\int D^{J\ast}_{M\lambda}(\phi,\theta,0) |\theta,\phi,\lambda\rangle d\Omega \\ &=\sum_{m_S}\sqrt{\frac{2J+1}{4\pi}}\int D^{J\ast}_{M\lambda}(\phi,\theta,0) |\theta,\phi\rangle D^1_{m_S\lambda}(\phi,\theta,0)|1,m_S\rangle d\Omega\\
&=\sum_{m_S,l,m}\sqrt{\frac{2l+1}{4\pi}}\sqrt{\frac{2J+1}{4\pi}}\int D^{J\ast}_{M\lambda}(\phi,\theta,0) |l,m\rangle D^l_{m0}(\phi,\theta,0) D^1_{m_S\lambda}(\phi,\theta,0)|1,m_S\rangle d\Omega\,.
\end{split}
\end{equation}
We use the following relation for the Wigner rotation functions
\begin{equation}
D^j_{mn}(R) D^{j^\prime}_{m^\prime n^\prime}(R) = \sum_{J,M,N} C(mm^\prime(jj^\prime)JM) \rangle D^J_{MN}(R) C(nn^\prime(jj^\prime)JN)\,,
\end{equation}
which yields
\begin{equation}\label{coneq2}
\begin{split}
|J,M,\lambda\rangle &=\sum_{m_S,l,m,l^\prime}\sqrt{\frac{2l+1}{4\pi}}\sqrt{\frac{2J+1}{4\pi}}\int D^{J\ast}_{M\lambda}(\phi,\theta,0) D^{l^\prime}_{m+m_S,\lambda}(\phi,\theta,0)d\Omega |l,m\rangle |m1\rangle\\
&\cdot C(m_Sm (l1)l^\prime m_S+m) C(0\lambda(l1)l^\prime\lambda)\\
&= \sum_{m_S,l,m,l^\prime}\sqrt{\frac{2l+1}{4\pi}}\sqrt{\frac{2J+1}{4\pi}}2\pi\delta_{M,m+m_S}\int d^J_{M\lambda}(x) d^{l^\prime}_{M\lambda}(x)dx |l,m\rangle |1,m_S\rangle\\
&\cdot C(m_S m(l1)l^\prime m_s+m) C(0\lambda(l1)l^\prime\lambda) \\
&= \sum_{l,m_S}\sqrt{\frac{2l+1}{2J+1}} C(m_S (M-m_S) (l1)J M)C(0\lambda(l1)J\lambda)|l,M-m_S\rangle |1,m_S\rangle\,.
\end{split}
\end{equation}
Therefore, we get from Eq.(\ref{coneq1}) and Eq.(\ref{coneq2})
\begin{equation}
\langle J,M;l,1 |J,M,\lambda\rangle = \sqrt{\frac{2l+1}{2J+1}} C(0\lambda(l1)J\lambda)\,,
\end{equation}
where we used 
\begin{equation}
\sum_{m_S,m_{s^\prime}}C(m_S (M-m_S)(l1)J M)C((M^\prime-m_{s^\prime})m_{s^\prime}(l1)J^\prime M^\prime) = \delta_{JJ^\prime}\delta_{MM^\prime}\,.
\end{equation}

Now we will connect the helicity projectors to angular momentum projectors. In order to do so we notice that
\begin{equation}\label{mla}
\begin{split}
M_{\lambda\overline\lambda}&=\langle J,M,\overline\lambda|T|J,M,\lambda\rangle = \sum_{l,l^\prime}\langle J,M,\overline\lambda|J,M;l^\prime,1\rangle \langle J,M;l^\prime,1|T| J,M;l,1 \rangle\langle J,M;l,1| J,M,\lambda\rangle \\
&= \sum_{l,l^\prime}\langle J,M;l^\prime,1|T| J,M;l,1 \rangle \frac{\sqrt{(2l+1)(2l^\prime+1)}}{2J+1} C(0\lambda(l1)J\lambda) C(0\overline\lambda(l^\prime1)J\overline\lambda)\,,
\end{split}
\end{equation}
where we used that also the orbital angular momentum states build a complete basis. By building quotients of the respective amplitudes, we can pin down constraints. If we only consider $J^P=1^+$ and therefore only deal with $s$- and $d$-waves, we know for all possible combinations of $l$
\begin{equation}
\frac{M_{11}}{M_{1-1}}=\frac{M_{11}}{M_{-1-1}}=\frac{M_{11}}{M_{-11}}=\frac{M_{10}}{M_{-10}}=\frac{M_{01}}{M_{0-1}}=1\,.
\end{equation}
Therefore, we can use
\begin{align}
M_{11}^+ &= M_{11} + M_{1-1} = 2 M_{11}\\
M_{10}^+ &= \frac{1}{\sqrt 2}(M_{10} + M_{-10})=\sqrt 2 M_{10}\\
M_{01}^+ &= \frac{1}{\sqrt 2}(M_{01} + M_{0-1})=\sqrt 2 M_{01}\,.
\end{align}
Looking up the Clebsch-Gordan coefficients of Eq.(\ref{mla}), we get the following relations for the respective transitions
\begin{align}
s-\text{wave}\rightarrow s-\text{wave}&:\qquad \frac{M_{11}^+}{M_{10}^+} = \sqrt 2\phantom{-}\,,\qquad \frac{M_{11}^+}{M_{01}^+} = \sqrt 2 \phantom{-}\,,\qquad \frac{M_{11}^+}{M_{00}} = 2 \\
s-\text{wave}\rightarrow d-\text{wave}&:\qquad \frac{M_{11}^+}{M_{10}^+} = -\frac{\sqrt 2}{2}\,,\qquad \frac{M_{11}^+}{M_{01}^+} = \sqrt 2\phantom{-}\,,\qquad \frac{M_{11}^+}{M_{00}} = - 1 \\
d-\text{wave}\rightarrow s-\text{wave}&:\qquad \frac{M_{11}^+}{M_{10}^+} = \sqrt 2\phantom{-}\,,\qquad \frac{M_{11}^+}{M_{01}^+} = -\frac{\sqrt 2}{ 2}\,,\qquad \frac{M_{11}^+}{M_{00}} = -1 \\
d-\text{wave}\rightarrow d-\text{wave}&:\qquad \frac{M_{11}^+}{M_{10}^+} = -\frac{\sqrt 2}{ 2}\,,\qquad \frac{M_{11}^+}{M_{01}^+} = -\frac{\sqrt 2}{ 2}\,,\qquad \frac{M_{11}^+}{M_{00}} = \frac{1}{2} \,.
\end{align}
Calling the transitions with definite angular momentum $D_{ab}$, where $a,b \in \lbrace s,d \rbrace$ and supressing the Lorentz indices, we get
\begin{align}
D_{ss} &= Y_{11} + \frac{1}{\sqrt 2} Y_{10} + \frac{1}{\sqrt 2} Y_{01} + \frac{1}{2} Y_{00}\,,\\
D_{sd} &= Y_{11} - \frac{2}{\sqrt 2} Y_{10} + \frac{1}{\sqrt 2} Y_{01} -  Y_{00}\,,\\
D_{ds} &= Y_{11} + \frac{1}{\sqrt 2} Y_{10} - \frac{2}{\sqrt 2} Y_{01} -  Y_{00}\,,\\
D_{dd} &= Y_{11} - \frac{2}{\sqrt 2} Y_{10} - \frac{2}{\sqrt 2} Y_{01} + 2 Y_{00}\,.
\end{align}
In principle we can multiply each of these expressions by an arbitrary normalisation constant, which we choose to be one, which means we use the above expressions. If we want to express our amplitude in terms of orbital angular momentum
\begin{equation}
\begin{split}
T^{\mu\nu} &= M_{11}^+ Y_{11}^{\mu\nu} + M_{10}^+ Y_{10}^{\mu\nu} + M_{01}^+ Y_{01}^{\mu\nu} + M_{00}^+ Y_{00}^{\mu\nu}\\
&= c_{ss} D_{ss}^{\mu\nu} + c_{sd} D_{sd}^{\mu\nu} + c_{ds} D_{ds}^{\mu\nu} + c_{dd} D_{dd}^{\mu\nu}\,,
\end{split}
\end{equation}
we have to solve the following equations
\begin{equation}
\begin{pmatrix}
1 & 1 & 1 & 1 \\
\frac{1}{\sqrt 2} & -\frac{2}{\sqrt 2} & \frac{1}{\sqrt 2} & -\frac{2}{\sqrt 2}\\
\frac{1}{\sqrt 2} & \frac{1}{\sqrt 2} & -\frac{2}{\sqrt 2}& -\frac{2}{\sqrt 2}\\
\frac{1}{2} & -1 & -1 & 2
\end{pmatrix}\cdot
\begin{pmatrix}
c_{ss}\\
c_{sd}\\
c_{ds}\\
c_{dd}
\end{pmatrix}=
\begin{pmatrix}
M_{11}^+\\
M_{10}^+\\
M_{01}^+\\
M_{00}
\end{pmatrix}\,.
\end{equation}
The solution to that equation is
\begin{equation}\label{ces}
\frac{1}{9}\begin{pmatrix}
4 & 2 \sqrt 2 & 2 \sqrt 2 & 2 \\
2 & -2\sqrt 2 & \sqrt 2 & -2\\
2 & \sqrt 2 & - 2 \sqrt 2 & - 2\\
1 & -\sqrt 2 & -\sqrt 2 & 2
\end{pmatrix}\cdot
\begin{pmatrix}
M_{11}^+\\
M_{10}^+\\
M_{01}^+\\
M_{00}
\end{pmatrix}=
\begin{pmatrix}
c_{ss}\\
c_{sd}\\
c_{ds}\\
c_{dd}
\end{pmatrix}\,.
\end{equation}
It is interesting to note, that only for a particle at rest, the Weinberg-Tomozawa term is a pure $s$-wave, while for moving particles, factors of $\frac{\omega}{M}$ reduce the $s$-wave part.

\section{Details on the $a_1$ calculation}\label{secdetails}

In this appendix we explicitly give the coefficients $A_i$ form Section \ref{a1inc}, which incorporate the nontrivial part of the calculation including the $a_1$. The explicit expressions read
\begin{equation}
A_1 = -\frac{f_A g V_{ud} g_V s}{\sqrt 2 F_0^3}(c_1 (M_1^{a1})_1 + c_2 (M_2^{a1})_1) + \sum_{\phi V} \frac{g V_{ud} g_V}{2 \sqrt 2 F_0^3}c_{\phi V} J_{\phi V}(\mu_2)\bigl(g_V \alpha_1^{\phi V} +\frac{1}{2}(f_V-2g_V)\alpha_2^{\phi V}\bigr)
\end{equation}
and
\begin{equation}
\begin{split}
A_2 &= -\frac{f_A g V_{ud} g_V s}{\sqrt 2 F_0^3} \frac{1}{\sqrt s p_{cm\pi\rho}^2} (c_1 (\omega_{\pi\rho}(M_1^{a1})_1 - \sqrt 2 M_\rho (M_1^{a1})_3) + c_2( \omega_{\pi\rho}(M_2^{a1})_1 - \sqrt 2 M_\rho(M_2^{a1})_3)) \\ 
&+ \sum_{\phi V} \frac{g V_{ud} g_V}{2 \sqrt 2 F_0^3}c_{\phi V}J_{\phi V}(\mu_2)\bigl(g_V \alpha_3^{\phi V} +\frac{1}{2}(f_V-2g_V)\alpha_4^{\phi V}\bigr)\,,
\end{split}
\end{equation}
where $c_{\pi\rho}=\sqrt 2$ and $c_{KK^\ast}=-1$. The first part of the coefficients $A_i$ contains the functions $M_i^{a1}$, which result from adding the last two diagrams of Fig. \ref{diags2} 
\begin{equation}
M^{a1}_1=\frac{1}{s-M_{a1}^2}(1+VJ)^{-1}
\begin{pmatrix}
\sqrt 2(s-m_\pi^2-M_\rho^2)\\
-(s-M_K^2-M_{K^\ast}^2)\\
\frac{M_\rho}{\sqrt s}(s+m_\pi^2-M_\rho^2)\\
-\frac{M_{K^\ast}}{\sqrt 2\sqrt s}(s+M_K^2-M_{K^\ast}^2)
\end{pmatrix} \,,
\end{equation}
\begin{equation}
M_2^{a1}=\frac{1}{s-M_{a1}^2}(1+VJ)^{-1}
\begin{pmatrix}
\sqrt 2(s+m_\pi^2-M_\rho^2)\\
-(s+M_K^2-M_{K^\ast}^2)\\
\frac{\sqrt s}{M_\rho}(s-m_\pi^2-M_\rho^2)\\
-\frac{\sqrt s}{\sqrt 2M_{K^\ast}}(s-M_K^2-M_{K^\ast}^2)
\end{pmatrix}\,,
\end{equation}
where $VJ$ is the matrix resulting from Eq.(\ref{bsalg}) and is given by
\begin{equation}
VJ=\begin{pmatrix}
V_{1111}J_1 & V_{1211} J_2 & V_{1110}J_1 & V_{1210} J_2 \\
V_{2111}J_1 & V_{2211} J_2 & V_{2110}J_1 & V_{2210} J_2 \\
V_{1101}J_1 & V_{1201} J_2 & V_{1100}J_1 & V_{1200} J_2 \\
V_{2101}J_1 & V_{2201} J_2 & V_{2100}J_1 & V_{2200} J_2 
\end{pmatrix}\,.
\end{equation}
The remaining part of the coefficients $A_i$ corresponds to the diagrams Fig. \ref{diags2}e, \ref{diags2}f. The $\alpha_i$ contain the expansion coefficients of the scattering amplitude $M_{abij}$ and are given by
\begin{equation}
\alpha_1^{\pi\rho} = 2 M_\rho^2 M_{1111} + \sqrt 2 \omega_{\pi\rho} M_\rho M_{1112}\,, \quad \alpha_1^{KK^\ast} = 2 M_{K^\ast}^2 M_{1211} + \sqrt 2 \omega_{KK^\ast} M_{K^\ast} M_{1212}\,,
\end{equation}
\begin{equation}
\alpha_2^{\pi\rho} = (s - m_\pi^2 + M_\rho^2) M_{1111} + \sqrt 2 \sqrt s M_\rho M_{1112}\,,\quad \alpha_2^{KK^\ast} = (s - m_\pi^2 + M_{K^\ast}^2) M_{1211} + \sqrt 2 \sqrt s M_{K^\ast} M_{1212}\,,
\end{equation}
\begin{align}
\alpha_3^{\pi\rho} &= \frac{1}{p^2_{cm\pi\rho} \sqrt s}(\omega_{\pi\rho} \alpha_1^{\pi\rho} - \sqrt 2 M_\rho( 2M_\rho^2 M_{1121} + \sqrt 2 M_\rho \omega_{\pi\rho} M_{1122})\,,\\
\alpha_3^{KK^\ast} &= \frac{1}{p^2_{cmKK^\ast} \sqrt s}(\omega_{KK^\ast} \alpha_1^{KK^\ast} - \sqrt 2 M_\rho( 2M_{K^\ast}^2 M_{1221} + \sqrt 2 M_{K^\ast} \omega_{KK^\ast} M_{1222})\,,
\end{align}
\begin{align}
\alpha_4^{\pi\rho} &= \frac{1}{p_{cm\pi\rho}^2 \sqrt s}(\omega_{\pi\rho} \alpha_2^{\pi\rho} - 2 s M_\rho (\sqrt 2 \omega_{\pi\rho} M_{1121} + M_\rho M_{1122}))\,,\\
\alpha_4^{KK^\ast} &= \frac{1}{p_{cmKK^\ast}^2 \sqrt s}(\omega_{KK^\ast} \alpha_2^{KK^\ast} - 2 s M_{K^\ast} (\sqrt 2 \omega_{KK^\ast} M_{1221} + M_{K^\ast} M_{1222}))\,.
\end{align}

\bibliography{litfilep}

\end{document}